\documentclass[a4paper,11pt]{article}

\usepackage{jcappub} 
\usepackage[T1]{fontenc} 
\usepackage[utf8]{inputenc}
\usepackage{xcolor}
\usepackage{soul}
\usepackage{cleveref}

\newcommand{\es}[2] {\begin{equation} \label{#1} \begin{split} #2 \end{split} \end{equation}}
\newcommand{\D}{{\rm d}}
\newcommand{\omc}{C_{\ell,\rm{GW}}^c (\bar{\Omega}_{\rm GW}^c)^2}

\newcommand{\clc}{C^c_{\ell,{\rm GW}}}
\newcommand{\cla}{C^a_{\ell,{\rm GW}}}

\def\lsim{\mathrel{\rlap{\lower4pt\hbox{\hskip1pt$\sim$}}
     \raise1pt\hbox{$<$}}}         
\def\gsim{\mathrel{\rlap{\lower4pt\hbox{\hskip1pt$\sim$}}
     \raise1pt\hbox{$>$}}}         
\title{Unraveling Cosmological Anisotropies within Stochastic Gravitational Wave Backgrounds}
\author[a]{Yanou Cui,}
\author[b,c]{Soubhik Kumar,}
\author[d]{Raman Sundrum,}
\author[e]{and Yuhsin Tsai}
\affiliation[a]{Department of Physics and Astronomy, University of California, Riverside, CA 92521, USA}
\affiliation[b]{Berkeley Center for Theoretical Physics, Department of Physics, \\University of California, Berkeley, CA 94720, USA}
\affiliation[c]{Theoretical Physics Group, Lawrence Berkeley National Laboratory, Berkeley, CA 94720, USA}
\affiliation[d]{Maryland Center for Fundamental Physics, Department of Physics, \\University of Maryland, College Park, MD 20742, USA}
\affiliation[e]{Department of Physics, University of Notre Dame, IN 46556, USA}
\emailAdd{yanou.cui@ucr.edu}
\emailAdd{soubhik@berkeley.edu}
\emailAdd{raman@umd.edu}
\emailAdd{ytsai3@nd.edu}

\begin{document}
\abstract{
Identifying the anisotropies in a cosmologically sourced stochastic gravitational wave background (SGWB) would be of significance in shedding light on the nature of primordial inhomogeneities. 
For example, if SGWB carries isocurvature fluctuations, it would provide evidence for a multi-field inflationary origin of these inhomogeneities.
However, this is challenging in practice due to finite detector sensitivity and also the presence of the astrophysical foregrounds that can compete with the cosmological signal.
In this work, we explore the prospects for measuring cosmological SGWB anisotropies in the presence of an astrophysical counterpart and detector noise.
To illustrate the main idea, we perform a Fisher analysis using a well-motivated cosmological SGWB template corresponding to a first order  phase transition,
and an astrophysical SGWB template corresponding to extra-galactic binary mergers, and compute the uncertainty with which various parameters characterizing the isotropic and anisotropic components can be extracted. 
We also discuss some subtleties and caveats involving shot noise in the astrophysical foreground.
Overall, we show that upcoming experiments, e.g., LISA, Taiji, Einstein Telescope, Cosmic Explorer, and BBO, can all be effective in discovering plausible anisotropic cosmological SGWBs.
}
\maketitle
\section{Introduction}
With the discovery of the binary black hole merger by the LIGO/Virgo collaboration in 2015~\cite{LIGOScientific:2016aoc}, gravitational waves (GW) have become a powerful new long-range messenger of our Universe.
Since then many such mergers, also involving neutron stars~\cite{LIGOScientific:2017ync}, have been discovered, ushering in a new era of multi-messenger astronomy.
Very excitingly, pulsar timing array (PTA) measurements have recently reported~\cite{NANOGrav:2023gor, Antoniadis:2023ott} evidence for, not an individual merger, but rather a stochastic gravitational wave background (SGWB). 
Such an SGWB might be formed by the combination of numerous, far-away super-massive binary black hole inspirals in nano-Hz frequency ranges or it might have a cosmological origin~\cite{NANOGrav:2023hvm, Antoniadis:2023xlr} (see Refs.~\cite{Christensen:2018iqi, Renzini:2022alw, vanRemortel:2022fkb} for recent reviews on SGWB).
A similar detection in $\sim$Hz frequencies could also be around the corner, especially with Advanced LIGO runs, with the $A^+$ upgrade~\cite{LIGOAplus}.
With confirmation of such a discovery of SGWB, a natural next step would be to characterize its strength and properties.
In particular, {\it anisotropies} of an SGWB can give us extremely valuable information, as we explore in this work.

Cosmological SGWB, as an observable, can be enormously useful from the perspective of probing particle physics and the very early Universe.
This is especially because following their production GW can propagate almost freely, and therefore bring pristine information from the primordial epoch.
Since various plausible sources of GW could have been active before the decoupling of the cosmic microwave background (CMB) photons or big bang nucleosynthesis (BBN), GW might be a unique observable using which we can probe the pre-BBN era, i.e., the primordial dark age~\cite{Boyle:2005se,Boyle:2007zx}.
Such early Universe sources include cosmic phase transitions (PT), cosmic strings, preheating after inflation, and enhanced scalar curvature perturbations at small scales among others (for reviews on cosmological sources of SGWB see, e.g., Refs.~\cite{Caprini:2018mtu, Caldwell:2022qsj}).

The recent evidence for an SGWB~\cite{NANOGrav:2023gor, Antoniadis:2023ott} is an intriguing milestone.
At the same time, some of the most plausible cosmological sources of SGWB, such as a PT or preheating, predict {\it anisotropies} within the associated SGWB~\cite{Geller:2018mwu, Bethke:2013aba, Bethke:2013vca}.\footnote{
Anisotropic SGWB could also arise if primordial fluctuations of scalar and tensor modes are non-Gaussian, see, e.g.,~\cite{Dimastrogiovanni:2019bfl, Adshead:2020bji, Dimastrogiovanni:2021mfs}.}
In the case of a PT, these anisotropies would be present since the sector undergoing PT would typically obtain inflationary large-scale density fluctuations\footnote{In principle, the very inhomogeneous PT sector itself creates anisotropies at length scales $1/H_{\rm PT}$, where $H_{\rm PT}$ is the Hubble scale during PT. 
However, these will be at extremely small and unresolvable angular scales when compared with the present-day Hubble scale.
Therefore, our focus is on pre-existing, inflationary superhorizon perturbations in the PT sector.}
which would get imprinted on the resulting SGWB when the PT takes place.
An observation of these anisotropies would provide us with a new map of the primordial, inhomogeneous Universe, potentially different from the CMB.
Thus a non-trivial cross-correlation (or lack thereof) between CMB and SGWB anisotropies, can reveal the existence of new degrees of freedom or isocurvature perturbations in the early Universe~\cite{Geller:2018mwu, Kumar:2021ffi, Bodas:2022zca, Bodas:2022urf}.
In fact, even without cross-correlating, if we confirm a cosmological origin of an SGWB and find that anisotropies of such an SGWB are much larger than $10^{-5}$, that alone would provide strong evidence that isocurvature fluctuations are generated in the early Universe.
A discovery of isocurvature perturbations would be very significant since it may hint toward multi-field dynamics during inflation.
It has been noted in Ref.~\cite{Geller:2018mwu} that future detectors such as LISA, DECIGO, and BBO can have a sufficient angular resolution so as to detect such anisotropies, at least for PT-generated SGWB.
Anisotropies in SGWB would also get induced via other means, for example, during the propagation of GW due to effects analogous to the Sachs-Wolfe and integrated Sachs-Wolfe effects~\cite{Contaldi:2016koz, Bartolo:2019yeu, Bartolo:2019oiq} within the CMB anisotropies. 
In addition, anisotropic SGWB can arise at relatively late times, for example, in the presence of cosmic strings~\cite{Olmez:2011cg, Kuroyanagi:2016ugi, Jenkins:2018nty, Cai:2021dgx}.

A crucial challenge in extracting primordial physics from an SGWB and its anisotropies comes from the fact that astrophysical sources, such as individually unresolvable binaries, would also give rise to an anisotropic SGWB.
Such anisotropies arise as the galaxies hosting the mergers themselves follow an inhomogeneous distribution.
Therefore, it is important to develop methodologies that can distinguish between the two classes of anisotropies.
In this context, we will denote astrophysical and cosmological anisotropies as $A_a$ and $A_c$, respectively.
To explain the notation, which will be further detailed in Sec.~\ref{sec:sources}, $A_c$ roughly captures the intensity fluctuation of a cosmological SGWB.
That is, an adiabatic $ {\cal O} (10^{-5})$ primordial fluctuation on top of an isotropic GW abundance (as a fraction of critical density) $\Omega_{\rm GW} \sim 10^{-8}$, for example, would correspond to $A_c \sim 10^{-5} \times 10^{-8}$.
The same notation is also followed for $A_a$. 

Existing work has focused on distinguishing between several components of SGWB, but primarily focusing on the isotropic component \cite{Parida:2015fma, Karnesis:2019mph, Caprini:2019pxz, Biscoveanu:2020gds, Flauger:2020qyi, Pieroni:2020rob, Barish:2020vmy, Martinovic:2020hru, Boileau:2021sni, Boileau:2021gbr, Gowling:2021gcy, Lewicki:2021kmu, Boileau:2022ter, Boileau:2022uos, Racco:2022bwj, Baghi:2023qnq} (see also~\cite{Mukherjee:2019oma, Mukherjee:2020jxa} for ideas utilizing the time-dependence of an SGWB).
An important feature that has been used so far is that the astrophysical and cosmological components typically have different frequency dependencies.
In this work, we build upon these ideas to further discriminate between {\it anisotropic} components.
We utilize the fact that the power spectrum of anisotropies from an astrophysical source would have different behavior as a function of multipole mode $\ell$, compared to that from a cosmological source.
This, coupled with the angular sensitivity of upcoming GW detectors, enables us to discriminate between the two sources.
We note that the extraction of cosmological anisotropies could help confirm the cosmological nature of the SGWB itself since both the monopole and the anisotropies would have identical frequency spectra, under certain reasonable assumptions that we will elaborate on in Sec.~\ref{sec:sources}. 

In detail, we carry out a Fisher analysis to understand the precision with which the cosmological and astrophysical sources can be distinguished.
We first note that because of the frequency- and $\ell$-dependencies of the astrophysical and cosmological signals, the parameters $A_a$ and $A_c$, that capture the magnitude of anisotropy at one fixed scale and frequency, are not sufficient to achieve a discrimination between the two sources. Therefore we need to parametrize both the frequency and $\ell$ dependence of the anisotropies of the two sources.
For the cosmological source, as an example, we consider a cosmological phase transition.
We parametrize the frequency dependence of the associated SGWB as a broken power law~\cite{Caprini:2009fx, Hindmarsh:2017gnf, Caprini:2019egz, Caprini:2018mtu, Ellis:2018mja, Ellis:2019oqb} dominated by bubble collisions.
We parametrize the frequency dependence of the astrophysical component as $f^{2/3}$, as is the case for merger-generated SGWB~\cite{Phinney:2001di, Farmer:2003pa}, assuming circular orbits.
For the cosmological anisotropies, we consider an approximately scale-invariant spectrum  $C_{\ell, \rm{GW}}^c \propto \ell^\kappa/\ell^{2}$ with $|\kappa| \ll 1$.
Astrophysical anisotropies were computed in several works~\cite{Cusin:2017fwz, Cusin:2018rsq, Jenkins:2018kxc, Cusin:2019jhg, Cusin:2019jpv, Bellomo:2021mer}.
In this article, we consider a dependence $C_{\ell,\rm{GW}}^a \propto \ell^\delta/\ell$ as derived in~\cite{Cusin:2017fwz, Cusin:2018rsq, Cusin:2019jhg, Cusin:2019jpv} with $|\delta|\ll 1$.
We note that while these correspond to motivated benchmark choices to illustrate our methodology, possibilities of other frequency and angular dependencies exist.
For example, if bubble collision, sound waves, and turbulence all contribute to an SGWB, the combined frequency dependence would be more involved than a broken power-law with just two spectral tilts.
Similarly, one can also consider a significant violation of scale invariance in the cosmological anisotropy component through certain `primordial features'~\cite{Chen:2010xka, Chluba:2015bqa} as studied in Ref.~\cite{Bodas:2022zca}.
For such scenarios, we expect our methodology can be adapted appropriately.

With these parametrizations at hand, we use the noise sensitivity as a function of $\ell$ ($\Omega_{\rm GW,n}^\ell$) of LISA, LISA-Taiji combination (LT), Einstein Telescope (ET), Cosmic Explorer (CE), and Big-Bang Observer (BBO) to obtain the $1\sigma$ forecast precision with which the anisotropic map can be reconstructed, as a function of $A_c$. 
For this purpose, we {\it assume} that the monopole of a cosmological SGWB has already been detected with its frequency dependence precisely measured.
Using that information, we then ask how well we can measure the anisotropies.\footnote{We note that this assumption can be relaxed and one can measure the frequency dependence from anisotropies separately if those are large enough.
A confirmation of the consistent frequency dependence of the monopole signal and anisotropies would then corroborate a specific origin of the associated SGWB.}
Therefore, we do not vary the cosmological frequency dependence in our Fisher analysis.
While we focus on these detectors, detection prospects for other detectors can also be obtained using similar methods given appropriate $\Omega_{\rm GW,n}^\ell$.
In particular, in this work, we restrict our attention to ground and space-based experiments in the mHz-Hz range.
A similar analysis could also be carried out in the context of PTA.

We now briefly summarize our findings in the context of these detectors.
For this purpose, we need to make some assumptions about the strength of both the monopole and the anisotropies of the astrophysical signal, a subject of active research.
Since our goal in the present work is to extract cosmological anisotropies in the presence of potential foregrounds, we take a conservative approach.
In particular, we assume the monopole of the (extra-galactic) astrophysical SGWB is as large as allowed by the current data~\cite{KAGRA:2021kbb}.\footnote{We discuss the contribution from the galactic white dwarf foreground in Sec.~\ref{sec:sources}.}
For lower frequencies, we extend this result assuming an $f^{2/3}$ scaling of $\Omega_{\rm GW}$.
For the anisotropies, we follow the Refs.~\cite{Cusin:2017fwz, Cusin:2018rsq, Cusin:2019jhg, Cusin:2019jpv}.
With these choices, we find the astrophysical anisotropies coming from extra-galactic binary mergers are likely unobservable at LISA and LT (more details can be found in Sec.~\ref{sec:sources}).
On the other hand, for an SGWB from a PT with $A_c \sim 10^{-11}$, LISA (LT) can observe some of the individual $\ell$-modes  at $\sim 10\%~(1\%)$ precision.
In Sec.~\ref{sec:sources}, we will give examples of how a large $A_c \sim 10^{-11}$ can arise while satisfying the current observational bounds on isocurvature perturbations~\cite{Planck:2018jri,Ghosh:2021axu}.
Restricting to adiabatic initial conditions instead, we find that a PT with a monopole strength of $\Omega_{\rm GW} \sim 10^{-8}$ (corresponding to PT parameters $\alpha \gtrsim 1$ and $\beta_{\rm PT}/H_{\rm PT} \sim 10$, reviewed below), LT would be able to observe up to $\ell \lesssim 5$ modes.

For frequencies relevant for ET/CE and BBO, both astrophysical and cosmological anisotropies are relevant.
While there are studies~\cite{Regimbau:2016ike, Sachdev:2020bkk} suggesting astrophysical foreground, especially from binary black holes, can be efficiently subtracted with the future ground-based detectors, recent LIGO data and related investigations~\cite{Zhou:2022nmt, Zhou:2022otw} found that such subtraction may be more challenging than previously thought and large residuals could remain.
Given this, we use the current upper limit on isotropic SGWB~\cite{KAGRA:2021kbb} for the astrophysical monopole signal strength, as above.
We then show that for $A_c \sim 10^{-11}$, for example, the combination of ET and CE would be able to extract an anisotropic cosmological SGWB at $\sim 5-30\%$ precision, depending on the precise $\ell$-mode, and up to $\ell \lesssim 6$.
For the same $A_c$, ET alone would be able to extract cosmological signals corresponding to $\ell=2$ and $\ell=4$ modes.

In the frequency band $\sim$~0.1~Hz--1~Hz, relevant for DECIGO or BBO, Ref.~\cite{Cutler:2005qq} showed that it might be possible to subtract binary mergers individually so as to reduce astrophysical foreground by a factor $\sim 10^{2}$.
However, again taking a conservative approach for the astrophysical monopole, we extend the current upper limit~\cite{KAGRA:2021kbb} to the 0.1~Hz--1~Hz frequency band, as above.
Even with these conservative choices, we will show that it is possible to extract cosmological anisotropy in the presence of astrophysical foreground in Sec.~\ref{sec:results}. 
As an example, for $A_c \sim 5\times 10^{-14}$, we find $A_c$ can be extracted with $\sim 20\%$ precision with BBO.
Larger $A_c$ for fixed $A_a$ would lead to better precision in measurements of $A_c$, as expected.
Similarly, if the astrophysical foreground can be subtracted, a given value of $A_c$ can be measured with better precision, and also smaller values of $A_c$ can be accessed (unless it is limited by detector noise).
We note that these conclusions are based on Fisher forecast alone, and therefore realistic detector effects could degrade the forecast precision that we find.
We also clarify that we conservatively restrict our attention to $\ell\leq 6$ modes, since LISA, ET, CE would be less sensitive to higher $\ell$ modes.
However, we expect BBO to have powerful sensitivity to $\ell>6$ modes as well (see, e.g.,~\cite{Braglia:2021fxn}), and the inclusion of those modes would improve the measurement prospects of both the astrophysical and cosmological anisotropies.

The rest of this article is organized as follows.
We describe the various properties of astrophysical and cosmological anisotropies in Sec.~\ref{sec:sources}, including a discussion of the shot noise, while establishing the benchmark values of $A_a$ and $A_c$ for different frequency ranges.
We then set up the Fisher analysis in Sec.~\ref{sec:results} and use the benchmarks to estimate the $1\sigma$ forecast precision on $A_a, A_c$, and $\ell$ dependence.
We conclude in Sec.~\ref{sec:conc}.

\section{Sources of GW Anisotropy}
\label{sec:sources}
We start with a review of some of the relevant quantities that can be used to describe anisotropies in SGWB, following the notation in Ref.~\cite{LISACosmologyWorkingGroup:2022kbp}.
We denote the spectral energy density of gravitational waves as,
\begin{align}
\Omega_{\rm GW}(f,\vec{x}) = \frac{1}{\rho_{c,0}}\frac{\D\rho_{\rm GW}(f,\vec{x})}{\D\ln f},    
\end{align}
where $\rho_{c,0}$ is the critical energy density of the Universe at the present day, $\vec{x}$ is the location of the GW detector, and $f$ is the frequency of the GW.
To characterize anisotropy, we define the variable $\omega_{\rm GW}(\vec{x}, f, \hat{n})$ via
\begin{align}
\Omega_{\rm GW}(f,\vec{x}) = \frac{1}{4\pi}\int \D^2\hat{n} \omega_{\rm GW}(\vec{x}, f, \hat{n}),    
\end{align}
where $\hat{n}$ is a unit vector pointing towards the sky.
The density perturbations are then defined as,
\begin{align}
\delta_{\rm GW}(\vec{x}, f, \hat{n}) =  \frac{\omega_{\rm GW}(\vec{x}, f, \hat{n}) - \bar{\Omega}_{\rm GW}(f)}{\bar{\Omega}_{\rm GW}(f)}.  
\end{align}
Here $\bar{\Omega}_{\rm GW}(f)$ is derived from $\Omega_{\rm GW}(f,\vec{x})$ by averaging over $\vec{x}$. To analyze $\delta_{\rm GW}(\vec{x}, f, \hat{n})$, we go to the harmonic space and write,
\begin{align}
\delta_{\rm GW}(\vec{x}, f, \hat{n}) = \sum_{\ell,m}  \delta_{{\rm GW},\ell m}(\vec{x}, f) Y_{\ell m}(\hat{n}).
\end{align}
Using the multipole moments $\delta_{{\rm GW},\ell m}$, we can define the expected power spectrum $C_{\ell, \rm GW}$ as,
\begin{align}
\langle \delta_{{\rm GW},\ell m}(\vec{x}, f) \delta^*_{{\rm GW},\ell' m'}(\vec{x}, f) \rangle = C_{\ell, \rm GW}(f)\delta_{\ell\ell'}\delta_{m m'}.  
\end{align}
We can characterize the power in a single $\ell$-mode (averaged over $m$-modes) by,
\begin{align}
\Omega_{\rm GW}^\ell(f) \equiv \sqrt{C_{\ell, \rm GW}} \bar{\Omega}_{\rm GW}(f),
\end{align}
where we have dropped the argument $\vec{x}$ assuming we have averaged over all such positions.
We have also assumed a `factorization' between the $\ell$ dependence and the $f$ dependence, such that the anisotropy power spectrum $C_{\ell, \rm GW}$ is independent of $f$. 
Such a factorization ansatz have been widely applied in literature (e.g. \cite{Ali-Haimoud:2020iyz, KAGRA:2021mth}).
Although few studies have been done to justify this assumption, a dedicated study on GW anisotropies from astrophysical sources demonstrated that the frequency-direction factorization is valid for the $f$ and $\ell$ range of interest in our work, while deviation may occur at regimes with $f>50$ Hz \cite{Cusin:2019jpv} and large $\ell>100$.
Given that all our analyses are focused on frequencies smaller than 50~Hz and we consider $\ell \leq 6$, we will assume this factorization of the astrophysical foreground for the rest of the article.

For the PT-generated cosmological source, the frequency dependence is primarily determined by the subhorizon physics during the PT (except for the low-frequency tail, determined by causality). On the other hand, the superhorizon $\ell$-dependence is determined by inflationary era fluctuations.
Therefore, we expect a factorization to be valid for the cosmological source as well.

In order to generalize the above expressions in the presence of an astrophysical ($a$) and a cosmological ($c$) component, we note that all our operations above have been linear in the perturbations. Hence we can write,
\begin{align}
\delta_{{\rm GW},\ell m}(f) = \delta_{{\rm GW},\ell m}^c(f) \frac{\bar{\Omega}^c_{\rm GW}}{\bar{\Omega}_{\rm GW}} + \delta_{{\rm GW},\ell m}^a(f) \frac{\bar{\Omega}^a_{\rm GW}}{\bar{\Omega}_{\rm GW}}.
\end{align}
Here we have defined,
\begin{align}
\delta_{\rm GW}^{c/a}( f, \hat{n}) =  \frac{\omega_{\rm GW}^{c/a}(f, \hat{n}) - \bar{\Omega}_{\rm GW}^{c/a}(f)}{\bar{\Omega}_{\rm GW}^{c/a}(f)},  
\end{align}
and the total signal is a sum of the cosmological and astrophysical components,
\begin{align}
\bar{\Omega}_{\rm GW} =  \bar{\Omega}_{\rm GW}^c + \bar{\Omega}_{\rm GW}^a. 
\end{align}
In terms of $C_\ell$ and assuming no correlation between cosmological and astrophysical signals (as elaborated upon below), for simplicity, we get 
\begin{align}
C_{\ell, \rm GW} =  C_{\ell, \rm GW}^c \left(\frac{\bar{\Omega}^c_{\rm GW}}{\bar{\Omega}_{\rm GW}}\right)^2 +  C_{\ell, \rm GW}^a \left(\frac{\bar{\Omega}^a_{\rm GW}}{\bar{\Omega}_{\rm GW}}\right)^2.
\end{align}
Therefore, the sum of cosmological and astrophysical contributions can be written as
\begin{align}\label{eq:om_gw_ell}
(\Omega_{\rm GW}^\ell)^2\bigg\rvert_{\rm signal} = C_{\ell, \rm GW}^c  (\bar{\Omega}^c_{\rm GW})^2 + C_{\ell, \rm GW}^a  (\bar{\Omega}^a_{\rm GW})^2. 
\end{align}

We note that the astrophysical and the cosmological anisotropies may be correlated for multiple reasons: (a) they would undergo a similar (integrated) Sachs-Wolfe effect, and (b) they can originate from the same adiabatic source of primordial perturbations.
However, as we will see in Sec.~\ref{sec:results}, for LISA/LT and CE/ET we will be interested in cases where cosmological anisotropies are comparable to or larger than the astrophysical ones.
In such scenarios including correlation would not change the result significantly.
On the other hand, for BBO we will be considering situations where cosmological anisotropies are much smaller than the astrophysical ones.
In these cases, including a correlation is expected to allow a better extraction of $A_c$ due to a linear sensitivity of $(\Omega_{\rm GW}^\ell)^2$ to $\delta_{\rm GW}^c$, as opposed to quadratic.
However, properly accounting for the primordial correlation as the matter fluctuations evolve in the late Universe is technically more involved and we will leave it for future work.
On the other hand, when SGWB carries isocurvature fluctuations, we expect the correlations between astrophysical and cosmological anisotropies to be of minimal advantage.
In summary, our current assumption of no correlation provides a more conservative but simpler overall analysis.

Motivated by the relation in Eq.~\eqref{eq:om_gw_ell}, we now also include detector noise (assuming no correlation with the signal), 
\begin{align}\label{eq:total_aniso}
(\Omega_{\rm GW}^\ell)^2  = C_{\ell, \rm GW}^c  (\bar{\Omega}^c_{\rm GW})^2 + C_{\ell, \rm GW}^a  (\bar{\Omega}^a_{\rm GW})^2 + (\Omega_{\rm GW, n}^\ell)^2.
\end{align}
Here we have used the notation of Ref.~\cite{LISACosmologyWorkingGroup:2022kbp} to include the detector noise contribution.
The angular dependence of the cosmological and astrophysical components can be parametrized (given frequency-direction factorization, as discussed above) as,
\begin{align}\label{eq:Ac}
   C^c_{\ell, \rm GW} (\bar{\Omega}^c_{\rm GW})^2 = A_c^2 {\cal S}(f)^2\frac{\ell^{\kappa}}{\ell(\ell+1)},\\
    \label{eq:Aa}
     C^a_{\ell, \rm GW} (\bar{\Omega}^a_{\rm GW})^2 = A_a^2 \left(\frac{f}{f_0}\right)^{2\gamma} \frac{\ell^{\delta}}{(2\ell+1)}.
\end{align} 
For an approximately scale-invariant primordial cosmological spectrum, we expect $\kappa \approx 0$ at large scales, while  
the expected
astrophysical contribution corresponds to $\delta \approx 0$~\cite{Cusin:2019jpv}.
We will therefore set $\kappa = \delta = 0$ in the Fisher analysis, and then compute with what precision we can extract bounds on $|\kappa|$ and $|\delta|$.
Here ${\cal S}(f)$ characterizes the frequency dependence of the cosmological signal, and we have assumed a power-law frequency dependence for the astrophysical signal.
We explain the various parameters above in more detail in the following.

\subsection{Cosmological Contribution}
\label{sec:cosmo_contri}
As mentioned in the Introduction, several cosmological sources can produce an SGWB within the reach of upcoming experiments such as LISA, Einstein Telescope (ET), Cosmic Explorer (CE), and BBO.  
Given that all these GW signals get produced and propagate in the inhomogeneous Universe, an SGWB must be anisotropic. 
The resulting spectrum $C^c_{\ell,{\rm GW}}$ can vary among different cosmological sources and is sensitive to the evolution of cosmological perturbations. 
Therefore, an anisotropic SGWB can be a precious probe of the early Universe for studying the pre-BBN Universe. 
In particular, through anisotropies, an SGWB is directly sensitive to fluctuations of light fields during inflation, beyond just the inflaton. Therefore, a discovery of isocurvature perturbations within the SGWB anisotropies would give strong evidence for multi-field inflationary dynamics.

In this work, we focus on an SGWB originating from a first order PT as an example to illustrate the prospect of identifying an anisotropic cosmological SGWB in the presence of an astrophysical foreground. 
We consider a scenario where the PT happens above temperature $T\sim$~TeV so that the horizon size at the time of the PT is much smaller than the angular resolution of the GW detectors. 
The detectors, therefore, only see a diffuse GW background from superposing GW from sources within a large number of  causally independent Hubble patches that went through the PT. 
The {\it local} temperature when the PT completes in each finite region of the sky is nearly identical and equals the bubble nucleation temperature $T_n$.
However, since the thermal history of each Hubble patch depends on the primordial fluctuation, each point on the $T=T_n$ surface has a varying distance in redshift from us. 
The variation of the redshift leads to a modulation of the GW energy density as a function of angle on the sky, and the resulting SGWB is therefore anisotropic.

\subsubsection{Frequency and the $\ell$-mode Dependence}
If the sector going through the PT carries approximately scale-invariant inflationary perturbations, analogous to those of the CMB, then $C_{\ell,{\rm GW}}^c\propto [\ell(\ell+1)]^{-1}$ for the larger angular scales $\ell\lsim 50$. The Sachs-Wolfe effect only gives an extra $\mathcal{O}(10)\%$ enhancement to the scale-invariant spectrum for lower $\ell$ modes~\cite{ValbusaDallArmi:2020ifo}. For $\ell\gsim 100$, the integrated Sachs-Wolfe (ISW) effect dominates the correction~\cite{ValbusaDallArmi:2020ifo,Braglia:2021fxn}.
Since GW is produced long before the CMB decoupling, the SGWB anisotropy experiences the changes in metric perturbations for a larger duration compared to CMB fluctuations.
Therefore, the early-ISW effect gives a larger enhancement of $C_{\ell,{\rm GW}}^c$ at higher $\ell$ modes. 

Besides the anisotropy from the cosmological perturbations, the peculiar motion of a GW detector also generates kinematic anisotropies due to Doppler effect~\cite{Jenkins:2018nty}.
The size of the kinematic anisotropies depends on the frequency spectrum $\bar{\Omega}_{\rm GW}^c(f)$, and the effect introduces an additional frequency dependence to the total anisotropic spectrum. 
However, given that the PT signals we consider are chosen as benchmarks to have monopole components well above the LISA, ET/CE, and BBO sensitivities, we can measure $\bar{\Omega}_{\rm GW}^c(f)$ precisely and calculate the kinematic anisotropies by knowing the detector motion. 
Therefore, it is plausible that the Doppler signal can be subtracted from the anisotropic background, and we do not consider such kinematic anisotropies in this work. 
For different sources of the GW production, such as a large primordial scalar perturbation with a peaked spectrum~\cite{Ananda:2006af, Baumann:2007zm} or cosmic strings that generate GW for an extended period of time~\cite{Jenkins:2018nty, Cai:2021dgx}, the $C_{\ell,{\rm GW}}^c$ can have different $\ell$ dependence. 
However, one can still apply our analysis to these cosmological signals by changing the spectral shape of $C_{\ell,{\rm GW}}^c$. 

Primarily three processes determine GW production from a first order PT: bubble wall collisions, plasma sound waves, and magneto-hydrodynamic (MHD) turbulence. 
Recent simulations indicate that the contribution from sound waves and turbulence can dominate over bubble wall collisions,  depending on the microphysical models.
For concreteness, however, we will focus on scenarios where bubble collisions dominate the GW production, as in the case of a supercooled PT, e.g.,~\cite{Konstandin:2011dr, Agashe:2019lhy, Ellis:2020nnr}, where the plasma of the PT sector gets diluted via supercooling. 
For such cases, we can approximate the frequency dependence of the cosmological signal by~\cite{Huber:2008hg} 
\begin{align}\label{eq:cosmo_freq}
\mathcal{S}(f) = \left(\frac{(a+b)f^a f_*^b}{a f^{a+b}+b f_*^{a+b}}\right)  
\end{align}
with $a \approx 3$, $b \approx 1$, and $f_*$ being the peak frequency of the spectrum.
Taking such models as benchmarks, we will use the spectrum in Eq.~\eqref{eq:cosmo_freq} as an example for the Fisher analysis.
On the other hand, if sound waves and/or turbulence contributions dominate, then our analysis can be straightforwardly modified to include the appropriate frequency dependence, by including parameters beyond $a$ and $b$ for frequency characterization.

\subsubsection{Amplitude of the Perturbation}
The strength of the GW signal primarily due to bubble collisions can be approximated in the thin-wall regime by~\cite{Huber:2008hg}
\begin{equation}
\Omega_{\rm GW}^{\rm peak}h^2 =  1.3\times10^{-6}\left(\frac{H_{\rm PT}}{\beta_{\rm PT}}\right)^2 \left(\frac{\alpha}{1+\alpha}\right)^2.
\label{eq:rhoGW}
\end{equation}
Here we have assumed that the bubble walls propagate at close to the speed of light and the effective number of degrees of freedom in the thermal bath $g_* \approx 100$.
The parameter $\alpha$ determines the ratio of the released vacuum energy density to the energy density of the surrounding plasma.
In the supercooling regime $\alpha\gg 1$.
The inverse duration of the PT is determined by $\beta_{\rm PT}\equiv d\ln\Gamma/dt\approx -4+T_n dS_b/dT\rvert_{T_n}$ where $\Gamma$ is the bubble nucleation rate per unit volume, and $S_b$ is the bounce action at $T_n$.
Finally, the Hubble rate at $T_n$ is given by $H_{\rm PT}$.
This bubble collision-generated spectrum peaks at
\begin{eqnarray}\label{eq:ftoday}
f_* &=& 0.04\,{\rm mHz}~
\left(\frac{H_{\rm PT}}{\beta_{\rm PT}}\right)^{-1} \left(\frac{T_n}{\rm TeV}\right).
\end{eqnarray}

Having discussed the monopole SGWB strength, we now move on to the anisotropic signal. If the SGWB carries a primordial density perturbation $\delta_{\rm GW}$, the anisotropy of the SGWB is
\begin{equation}
\delta\Omega_{\rm GW}^{\rm peak}h^2  \equiv \delta_{\rm GW}\Omega_{\rm GW}^{\rm peak}h^2.
\label{eq:rhoGW}
\end{equation}
In the scenario where the inflaton reheats all sectors, the density perturbation would be adiabatic and we would have $\delta_{\rm GW}\approx (4/3)\sqrt{A_s}$ on large scales, where  $A_s=2.1\times 10^{-9}$~\cite{Planck:2018vyg} is fixed by the normalization of the CMB power spectrum. 
In the scenario where the PT sector comes from the reheating of a `curvaton' field~\cite{Linde:1996gt, Lyth:2001nq, Enqvist:2001zp, Moroi:2001ct} that carries different quantum fluctuations than the inflaton, the SGWB picks up isocurvature fluctuations that allow $\delta_{\rm GW}\gg\sqrt{A_s}$~\cite{Geller:2018mwu,Kumar:2021ffi}. 
In the isocurvature scenario, the primary observational bound comes from bounds on dark radiation isocurvature~\cite{Ghosh:2021axu}, since the emitted GW behaves as dark radiation. 
Saturating this bound implies $\delta \Omega_{\rm GW}^{\rm peak}h^2 \lesssim 10^{-10}$.
However, the monopole perturbation $\Omega_{\rm GW}$ can now be smaller and get compensated by a larger $\delta_{\rm GW}>(4/3)\sqrt{A_s}$ while ensuring the isocurvature bound. Cosmological mechanisms leading to such scenarios have been recently constructed in Ref.~\cite{Bodas:2022urf}. In the Fisher analysis, we pick a range of $A_c$ (as defined in Eq.~\eqref{eq:Ac}) that can arise from either the adiabatic or isocurvature perturbations. 
We now consider these two possibilities in more detail.

\paragraph{Adiabatic perturbations.}
In this case, the size of SGWB density perturbation is around 
\begin{equation}
\delta\Omega_{\rm GW}^{\rm peak}h^2  = 8\times10^{-11}\left(\frac{H_{\rm PT}}{\beta_{\rm PT}}\right)^2 \left(\frac{\alpha}{1+\alpha}\right)^2.
\label{eq:rhoGW}
\end{equation}
Depending on the PT model, $(\beta_{\rm PT}/H_{\rm PT})^2$ can vary from $10$ to $10^{4}$.
The lower range of $(\beta_{\rm PT}/H_{\rm PT})^2$ can be achieved for a non-standard SGWB from a PT (see, e.g., Ref.~\cite{Agashe:2019lhy}),  while the upper range comes from matching $\Gamma\sim H_{\rm PT}^4$. 
By choosing $\alpha\gtrsim1$, we then get $\delta\Omega_{\rm GW}^{\rm peak}\sim 10^{-14}-10^{-11}$.
We will see in Sec.~\ref{sec:fisher} that LISA, LT, or ET+CE can observe anisotropies for $\delta\Omega_{\rm GW}^{\rm peak} \gtrsim 10^{-13}$.
For $(\beta_{\rm PT}/H_{\rm PT})^2 \gtrsim 10^3$ and/or $\alpha\ll 1$, anisotropies become smaller than this, and BBO would be the primary probe of such anisotropies, as we will demonstrate. 

\paragraph{Isocurvature perturbations.}
In the case of adiabatic perturbation, we have seen that LISA would be sensitive to cosmological SGWB anisotropies only for a sufficiently  strong SGWB. 
However, in the presence of primordial isocurvature perturbations, this need not be the case.
For example, we can consider $\delta_{\rm GW} \sim 10^{-3}$, $\beta_{\rm PT} / H_{\rm PT} \sim 30$, and $\alpha \sim 1$ leading to $\Omega_{\rm GW}^{\rm peak} \sim 10^{-9}$ and $\delta \Omega_{\rm GW}^{\rm peak}h^2 \sim 10^{-12}$. 

In the following Fisher analysis, we parametrize the cosmological signal as in Eq.~\eqref{eq:Ac}.
According to the discussion above, we choose benchmark numbers $\kappa=0$ and $A_c\sim 10^{-14}-10^{-9}$.
Our choice of $\kappa=0$ is motivated since we assumed SGWB carries the standard, approximately scale-invariant primordial spectrum.

\subsection{Astrophysical Contribution}
An SGWB also naturally arises due to astrophysical effects. 
For the frequency bands around the peak sensitivities of experiments such as LISA, BBO, ET or CE, the dominant foreground comes from the incoherent superposition of a large number of unresolved sources, including merging stellar-mass black holes, binary neutron stars, and white dwarves~\cite{LIGOScientific:2016fpe,Regimbau:2016ike,Mandic:2016lcn,Dvorkin:2016okx,Nakazato:2016nkj,Dvorkin:2016wac}. Therefore, the anisotropy of the astrophysical GW background depends on the formation of large-scale structure and the local astrophysics on sub-galactic scales. The fluctuations are significantly different compared to the above-mentioned cosmological signals.

\subsubsection{Frequency and the $\ell$-mode Dependence}
Parameters of the $\Lambda$CDM model are now precisely measured~\cite{Planck:2018vyg}, which determines the Universe's expansion history from the beginning of stellar activity until today. 
The cosmology of structure formation is also well understood at scales $\gtrsim 10$~Mpc, which mainly contributes to the modes with $\ell\lesssim 1000$, the regime we focus on for studying the SGWB anisotropy power spectrum $C_{\ell,{\rm GW}}$. 
Different from the comparatively short duration of  production of GW during a specific epoch, such as from a PT which is the cosmological benchmark we choose, GW from astrophysical sources is produced over an extended period. 
As a result, the $\ell$-dependence of $C_{\ell,{\rm GW}}^{a}$ is different from that of its counterpart sourced from a PT, where $C_{\ell,{\rm GW}}^c\propto [\ell(\ell+1)]^{-1}$, as mentioned above. 
After taking into account the growth of the matter density perturbation and integrating the signal over time, one finds that the anisotropic spectrum is roughly proportional to $C_{\ell,{\rm GW}}^{a}\propto (2\ell+1)^{-1}$~\cite{Cusin:2017fwz, Cusin:2018rsq, Cusin:2019jhg, Cusin:2019jpv}. 

As discussed in Sec.~\ref{sec:sources}, we assume frequency-direction factorization for the astrophysical anisotropies and parametrize them via Eq.~(\ref{eq:Aa}), where $\delta\approx 0$.
For $f<50$~Hz, the frequency spectrum in Eq.~(\ref{eq:Aa}) is well approximated by $\gamma\approx \frac{2}{3}$~\cite{Phinney:2001di, Farmer:2003pa}. 
As we will see, the difference in the shape of the frequency and anisotropic spectra of astrophysical sources as compared to the PT signal plays an essential role in distinguishing the cosmological signal from astrophysical foreground. 
Even if our simple template of the astrophysical foreground~\eqref{eq:Aa} does not apply to higher frequency signals, one can still perform a similar Fisher analysis with a more complex $C_{\ell,{\rm GW}}^{a}$ in $(f,\ell)$ space by including the GW emission from other astrophysical sources.

\subsubsection{Amplitude of the Perturbation} The amplitude of the power spectrum $A_a$ has a significant uncertainty due to sub-galactic scale physics~\cite{Cusin:2019jpv}. Since $A_a$ depends on the time evolution of both the large scale structure and the formation of merger systems, various sub-galactic scale physics such as the black hole formation process or the merger rate~\cite{Cusin:2019jpv,Jenkins:2018kxc}, can significantly modify $A_a$. 
 Despite affecting $A_a$, the shape of the $C_{\ell,{\rm GW}}^a$ spectrum is only mildly modified at $\mathcal{O}(10)\%$ level, for $\ell\lsim 10$~\cite{Cusin:2019jhg}.

Despite this uncertainty in $A_a$, just as for cosmic variance, we can use a benchmark model for the astrophysical anisotropies with the understanding that in reality they may vary due to sub-galactic uncertainties.
However, we expect the benchmark model to represent the rough size of astrophysical anisotropy for each spherical harmonic.
This will allow us to extract our best estimate for the precision we can obtain for the cosomological anisotropies from the Fisher analysis.

LIGO/Virgo observations have set an upper limit on the size of the isotropic energy density $\bar\Omega_{\rm GW}^a(f)<3.4\times 10^{-9}$ ($95\%$ C.~L.), for $\gamma=2/3$~\cite{KAGRA:2021kbb} and reference frequency $25$~Hz. They also consider a fiducial model for $\bar\Omega_{\rm GW}(f)$ utilizing the data of observed merger events, with a central value a factor of 2-3 below this upper limit.
Therefore, for our benchmark choice, we conservatively take into account the above-mentioned upper limit, as our goal is to demonstrate how well we can extract the cosmological signature in the presence of astrophysical foregrounds. 
The number translates to $\bar{\Omega}_{\rm GW}^a\approx 7\times 10^{-12}$ around the frequency of peak sensitivity of LISA $f_0=2.5\times 10^{-3}$~Hz.

The astrophysical power spectrum $C^a_{\ell, \rm GW}$ varies somewhat in the literature~\cite{Jenkins:2018kxc, Cusin:2019jhg, Cusin:2019jpv, Bellomo:2021mer}.
We use the computation of~\cite{Cusin:2019jhg, Cusin:2019jpv} for concreteness, and take $(\ell+1/2)C^a_{\ell, \rm GW} \approx 4\times 10^{-4}$.
With this choice, and saturating the upper limit on the monopole $\bar{\Omega}_{\rm GW}^a$~\cite{KAGRA:2021kbb}, we find $A_a \simeq 10^{-13}$.
We will see in Sec.~\ref{sec:results} that these astrophysical anisotropies are likely unobservable at LISA and LT.
Using the upper limit in~\cite{KAGRA:2021kbb} and using the $f^{2/3}$ scaling, we find $\bar\Omega_{\rm GW}^a(f) \leq 10^{-10}$ at $0.25$~Hz.
An estimate similar to the above then gives $A_a \sim 2\times 10^{-12}$. 
The resulting anisotropies are indeed observable in BBO, and therefore, they need to be accounted for when extracting a cosmological signal.
Finally, a similar analysis gives $\bar{\Omega}^a_{\rm GW}(f) \leq 7\times 10^{-10}$ around ET/CE peak sensitivity frequency $6$~Hz, which corresponds to $A_a \sim 2\times 10^{-11}$.
Such anisotropies are also observable in ET and CE.
More quantitative results will be given in Sec.~\ref{sec:results}.

\paragraph{Foreground subtraction.}
Although our analysis follows a conservative assumption by saturating $A_a$ utilizing the upper limit in~\cite{KAGRA:2021kbb}, it is possible that future experiments can model and subtract the foreground with enough computing power. 
For example, Ref.~\cite{Cutler:2005qq} shows that under certain assumptions of the detector sensitivity, BBO has a chance to identify neutron star merges up to redshift $z\approx 3.6$, which allows the subtraction of $\approx 99\%$ of the foreground. 
Given the above upper limit at frequencies relevant for BBO, it means after a $10^2$ reduction the astrophysical monopole is below $\bar{\Omega}^a_{\rm GW}<10^{-12}$.
For the anisotropies, it would imply a strength, $A_a \sim 2\times 10^{-14}$.
A less efficient subtraction capability would imply a larger $A_a$.
In this regard, Ref.~\cite{Zhou:2022nmt} shows that when subtracting merger signals in the ET/CE frequency range, the residual background from imperfect removal of resolved sources can be large and can limit the sensitivity to the cosmological signal. 
Therefore, it is still important to use the $f$ and the $\ell$ dependence in identifying the cosmological signal as we investigate in this work. 
Given these considerations, in the Fisher analysis, we take a conservative approach and will saturate the astrophysical monopole strength at the current upper limit~\cite{KAGRA:2021kbb}.
An efficient subtraction would typically imply better prospect for detection of cosmological anisotropies.
To give an example, we consider the limit where astrophysical anisotropies are much larger than cosmological anisotropies and detector noise, both before and after foreground subtraction.
In that case, a reduction in $\bar{\Omega}_{\rm GW}^a$ by a factor of $x$, through foreground subtraction, would imply the precision on cosmological anisotropies would roughly improve by the same factor of $x$.
This can be seen by considering the explicit form of the Fisher matrix in Eq.~\eqref{eq:fish_mat_f}.

\paragraph{Shot noise.}
The power spectrum of the astrophysical foreground contains shot noise components arising from finite sampling from the underlying galaxy distribution and average binary coalescence rate~\cite{Jenkins:2018kxc,Jenkins:2019uzp}. 
While the spatial shot noise, resulting from the discreteness of galaxy distribution, has negligible impact on $C_{\ell,{\rm GW}}^a$ for low $\ell$ modes ($<10$)~\cite{Cusin:2019jpv} that we consider, the temporal shot noise (also called popcorn noise) originating from discretized compact binary coalescence event rates can be significant. 
It can dominate over the discussed anisotropic foreground in the frequency band of ET and CE.

Proposed approaches, such as cross-correlation with  dense galaxy maps, have been suggested to mitigate both types of shot noise~\cite{Cusin:2019jpv,Alonso:2020mva}. 
Sensitivities from ET/CE measurements are expected to further reduce temporal shot noise and resolve additional popcorn events at a higher redshift~\cite{Alonso:2020mva}. 
However, without precise knowledge of the performance of ET/CE or the effectiveness of the proposed remedies, it is challenging to make a meaningful estimation at this stage. 
Therefore, in this study, we optimistically assume that these remedies successfully eliminate shot noise in the ET/CE measurements and omit them from the Fisher analysis.
For the LISA and BBO bands, however, the foreground originates from binary systems in the inspiralling phase, lasting much longer than the observation time span. Hence, the popcorn shot noise in these bands is absent~\cite{Cusin:2019jpv}, and only spatial shot noise is present. However, as mentioned above, for $\ell<10$, the spatial shot noise is subdominant to the `theoretical' astrophysical anisotropy power spectrum that we have been considering so far. 
Given that we do not include popcorn noise for ET/CE, in Sec.~\ref{sec:results}, we first discuss the projections for LISA/LT and BBO, and then move on to the comparatively less robust results for ET/CE.

\subsubsection{Signal from Other Astrophysical Sources} 
Before ending the discussion of astrophysical signals, we briefly comment on astrophysical sources other than the stellar-mass black holes and neutron stars which are expected to be the leading sources for our study. 
For instance, LISA is sensitive to GW emission from white dwarf (WD) mergers in the Milky Way, and we expect most of the WD binaries to be unresolved and produce a stochastic signal~\cite{Adams:2013qma,Korol:2021pun}. 
The white dwarf signal can be highly anisotropic and have an annual modulation due to detector motion.
The frequency spectrum of this galactic signal will not be a simple power law as well~\cite{Boileau:2020rpg,Korol:2021pun}. 

There have been proposals regarding separating the WD signal from the other astrophysical foregrounds and subtracting it, e.g.,~\cite{Thrane_2009, Adams:2013qma, Pieroni:2020rob}.
Furthermore, as estimated in, e.g., Ref.~\cite{Korol:2021pun}, the unresolved WD signals become significant at frequencies slightly below the peak sensitivity of LISA while rapidly declining for $f\gsim 3$~mHz.
On the other hand, our benchmark cosmological PT signals in the LISA/LT range have peak frequencies at  4~mHz and 10~mHz, and these choices ensure that the WD signal does not significantly interfere with our ability to isolate the cosmological signal, even without any subtraction.
Given these reasons, we do not consider white dwarf mergers in our analysis.

Another source, potentially relevant for LISA and mid-band detectors, originates from intermediate mass black holes (IMBHs), with masses between the stellar mass ($\mathcal{O}(10)\,M_{\odot}$) and supermassive black holes ($\mathcal{O}(10^4)\,M_{\odot}$) \cite{Amaro-Seoane:2007osp, Amaro-Seoane:2018gbb}. Although IMBH was hypothetical, the recent GW190521 event has provided indirect evidence for the existence of such black holes ~\cite{LIGOScientific:2020iuh}. One can then postulate Intermediate Mass Ratio Inspirals (IMRIs) resulting from the merger of stellar mass black holes and IMBHs. Ref.~\cite{Barish:2020vmy} estimates the $\bar\Omega_{{\rm GW}}(f)$ of the IMRIs using a similar model as for stellar mass binary black holes and modifying only the mass distribution of the IMBH and the fiducial merger rate. The estimate gives a stochastic GW background that is a bit smaller than signals from the stellar mass binaries merging near frequencies of the peak LISA and BBO sensitivities. 
The shape of frequency spectrum is also degenerate with the stellar mass mergers at frequencies below the highest frequency $\sim$$1$~Hz of the IMRI signal from the merger phase. Although more precise modeling is needed to better determine the IMRI signal, the result should not significantly change the signal spectrum assumed in Eq.~(\ref{eq:Ac}). 

Besides the IMRI, LISA is also sensitive to extreme mass ratio inspirals
(EMRIs), mergers between stellar mass and supermassive black holes \cite{Amaro-Seoane:2007osp, Babak:2017tow}. From the estimate in~\cite{Barish:2020vmy} based on \cite{Babak:2017tow, Bonetti:2020jku}, the EMRI signal can be comparable to the signal from the stellar mass mergers near the center of LISA frequency band. However, precise modeling of the overall signal from EMRIs is complex, and we expect a larger uncertainty in the estimate of EMRI signals than signals from stellar mass mergers. Given the current large uncertainties on IMRI and EMRI, we make the reasonable assumption that these sources are subdominant to the unresolved binary BH/NS sources and do not include them in the Fisher analysis.
With better understanding enabled by future data and further research, these other sources can be incorporated into the analysis in case observations prove their significance.

\section{Detector Sensitivity to an Anisotropic GW Sky}
\label{sec:results}
\subsection{Fisher Analysis}
To perform a Fisher analysis, we take into account the factorization between frequency and angular dependence.
 That is, as explained in Sec.~\ref{sec:sources}, we assume for the cosmological signal, each $\ell$-mode follows the same frequency dependence, and the same is assumed for the astrophysical signal as well.
Furthermore, as discussed before, we consider multipole modes up to $\ell\leq 6$,
and therefore cosmic variance plays an important role in determining the precision with which various fundamental parameters can be extracted.

Given these considerations, we carry out the Fisher analysis in two steps.
First, we ask how precisely a given cosmological and astrophysical $\ell$-mode can be separated purely based on the different frequency dependence of the cosmological and the astrophysical signal.
At this step, the parameters $\kappa$ and $\delta$, determining the spectral tilt in Eq.~\eqref{eq:Ac} and Eq.~\eqref{eq:Aa}, do not play any role since we analyze each $\ell$-mode separately.
Therefore, these results are model-independent and can be used to further study any angular dependence of the signals.
In the next step, we use the above results to perform another Fisher analysis to determine the precision on $\kappa$, assuming an $\ell$-dependence as in Eq.~\eqref{eq:Ac}.

\label{sec:fisher}
\subsubsection{Frequency Space Analysis}
The Fisher matrix for the frequency space analysis for each individual $\ell$-mode is given by~\cite{LISACosmologyWorkingGroup:2022kbp},
\begin{align}
\label{eq:fish_mat_f}
   \left[ F_{\ell}\right]_{\alpha\beta} = 2(2\ell + 1) T_{\rm obs} \int \D f  \frac{\frac{\partial}{\partial\lambda_\alpha}\left(\bar{\Omega}_{\rm GW}\sqrt{C_\ell}\right) \frac{\partial}{\partial\lambda_\beta}\left(\bar{\Omega}_{\rm GW}\sqrt{C_\ell}\right)}{C_\ell (\bar{\Omega}_{\rm GW})^2+\left(\Omega_{\rm GW,n}^\ell\right)^2}.
\end{align}
Here, we sum over all the $m$ modes for a given $\ell$-mode, subject to the conservative assumption of uncorrelated astrophysical and cosmological anisotropies,
\begin{align}
	C_\ell (\bar{\Omega}_{\rm GW})^2 = C_{\ell, {\rm GW}}^c (\bar{\Omega}_{\rm GW}^c)^2 + C_{\ell, {\rm GW}}^a (\bar{\Omega}_{\rm GW}^a)^2.
\end{align}
We take the observation time to be $T_{\rm obs}=3$~years.
The Fisher matrix $F_{\ell}$ is two-dimensional, with $\lambda_1 = \ln C_{\ell,\rm{GW}}^c$ and $\lambda_2 = \ln C_{\ell,\rm{GW}}^a$.
We also assume that the frequency dependencies of the cosmological (say, by measurement of a large monopole moment) and astrophysical signals, and the detector noise are known.
Then we need the following derivatives to evaluate the Fisher matrix,
\begin{align}
    \frac{\partial}{\partial\ln C_{\ell,\rm{GW}}^c}(\bar{\Omega}_{\rm GW}\sqrt{C_\ell}) = \frac{C_{\ell,\rm{GW}}^c (\bar{\Omega}_{\rm GW}^c)^2}{2\sqrt{C_{\ell,\rm{GW}}^c (\bar{\Omega}_{\rm GW}^c)^2 + C_{\ell,\rm{GW}}^a (\bar{\Omega}_{\rm GW}^a)^2}}
\end{align}
\begin{align}
    \frac{\partial}{\partial\ln C_{\ell,\rm{GW}}^a}(\bar{\Omega}_{\rm GW}\sqrt{C_\ell}) = \frac{C_{\ell,\rm{GW}}^a (\bar{\Omega}_{\rm GW}^a)^2}{2\sqrt{C_{\ell,\rm{GW}}^c (\bar{\Omega}_{\rm GW}^c)^2 + C_{\ell,\rm{GW}}^a (\bar{\Omega}_{\rm GW}^a)^2}}.
\end{align}
From the Fisher matrix, we can compute the fractional precision $\sigma$ with which the multipoles can be measured as $\sigma_{C_{\ell,\rm{GW}}^c}^2 = \left[F_\ell^{-1}\right]_{11}$ and $\sigma_{C_{\ell,\rm{GW}}^a}^2 = \left[F_\ell^{-1}\right]_{22}$. 

\subsubsection{Multipole Space Analysis}
Having used the frequency space information to separate the astrophysical and cosmological signals, we can analyze their $\ell$-space spectra separately.
This way we can determine the precision with which the spectral tilts $\kappa$ and $\delta$ in Eqs.~\eqref{eq:Ac} and~\eqref{eq:Aa} can be measured.
In particular, we focus on a benchmark $\kappa= 0$ and $\delta = 0$, and then compute the precision with which $\kappa$ and $\delta$ can be extracted.
The associated Fisher matrix, which is just a number in this case, for the cosmological component can be written as,
\begin{align}
F_c = \sum_\ell \frac{\left(\frac{\partial C^c_{\ell,\rm{GW}}}{\partial \kappa}\right) \left(\frac{\partial C^c_{\ell,\rm{GW}}}{\partial \kappa}\right)}{{\rm Var}(C_\ell)}
\end{align}
where ${\rm{Var}}(C_\ell) = \sigma_{C_{\ell,\rm{GW}}^c}^2 (C^c_{\ell,\rm{GW}})^2 + (C^c_{\ell,\rm{GW}})^2/(2\ell+1)$.
The precision $\sigma_{C_{\ell,\rm{GW}}^c}$ follows from the frequency space analysis, and the second contribution represents cosmic variance.
The above expression can be simplified,
\es{eq:sig_gamma}{
{1 \over \sigma_\kappa^2} = \sum_\ell {(\ln\ell)^2 \over {\sigma_{C_{\ell,\rm{GW}}^c}^2 + {1 \over 2\ell+1}}}.
}
The Fisher matrix $F_a$ for the astrophysical component to determine $\delta$ is defined analogously.
With these expressions, we can now evaluate the detection prospects of cosmological and astrophysical anisotropies at various GW detectors.
We organize this discussion based on the frequency ranges the detectors are sensitive to.

As discussed in Sec.~\ref{sec:sources}, both the spatial and temporal shot noise can be safely ignored for our analysis around the mHz band. 
Therefore, we start our discussion with LISA and LISA-Taiji.
For around 10~Hz frequencies, on the other hand, the temporal shot noise is significant.
While there are mitigation proposals~\cite{Cusin:2019jpv, Alonso:2020mva}, without a dedicated analysis in the context of ET/CE, it is not immediately clear to what extent the temporal shot noise can be mitigated.
As mentioned earlier, in this work, we take an optimistic approach by assuming that the temporal shot noise can be fully mitigated.
Given this assumption, we discuss our comparatively less robust results for ET/CE towards the end of this section.

\subsection{LISA and LISA-Taiji}
To forecast the detection prospects at LISA, we use {\tt schnell}~\cite{Alonso:2020rar} to obtain noise sensitivity curves $\Omega_{\rm GW,n}^\ell$ as a function of frequency.
We also specify parameters for Taiji and obtain $\Omega_{\rm GW,n}^\ell$ for the LISA-Taiji (LT) combination using {\tt schnell}.
We find that while LISA has the necessary sensitivity to only the $\ell=2,3,4$ modes, LT has powerful sensitivity to the other $\ell$ modes as well, $\ell=1$ through $\ell=6$. 
We will see below that these additional $\ell$-modes play an important role in constraining $\kappa$ and $\delta$, as expected. 

\begin{figure}
\begin{center}
\includegraphics[width=0.8\textwidth]{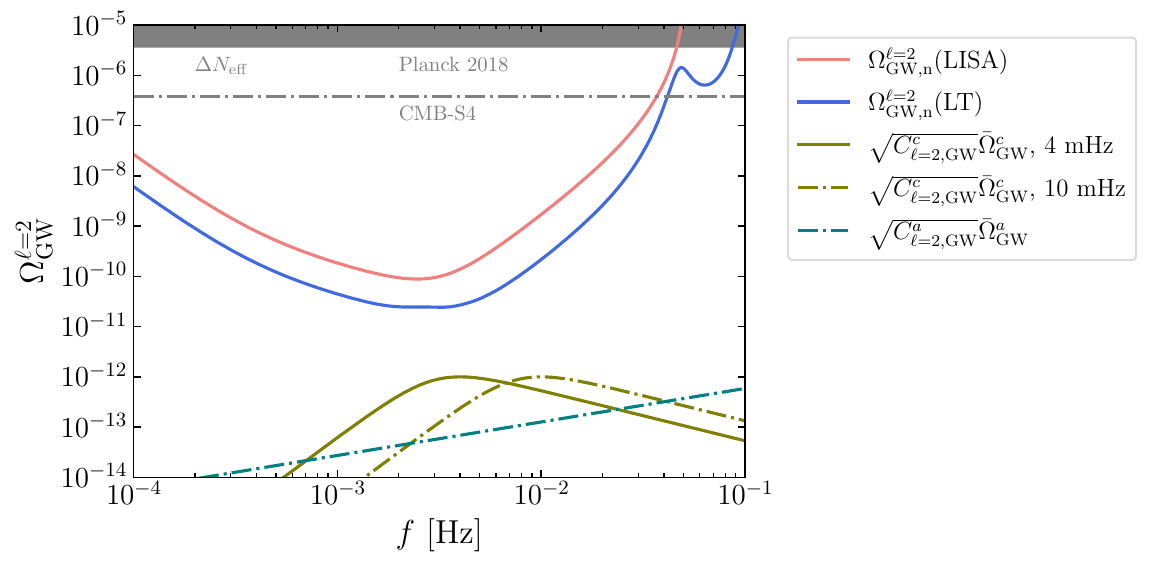}
\end{center}
\caption{Comparison between astrophysical anisotropy (dot-dashed teal), cosmological anisotropy (olive) with respect to detector sensitivity at LISA (red) and LT (blue) for the $\ell=2$ mode. We use $C_{\ell=2,\rm{GW}}^c = 10^{-8}$, $\bar{\Omega}^c_{\rm GW}\big\rvert_{\rm peak} = 10^{-8}$ with $f_* = 4$~mHz (solid olive) and $f_* = 10$~mHz (dot-dashed olive) for the cosmological signal. We also use $C_{\ell=2, \rm{GW}}^a = 2\times 10^{-4}$~\cite{Cusin:2019jhg, Cusin:2019jpv}. As explained in the text, since our purpose is to extract cosmological anisotropies, we conservatively saturate the astrophysical monopole to its current upper limit, $\bar{\Omega}_{\rm GW}^a = 10^{-9}$ at 10~Hz~\cite{KAGRA:2021kbb}, and then extend it to LISA/LT frequencies using a $f^{2/3}$ scaling, assuming circular binary orbits~\cite{Phinney:2001di, Farmer:2003pa} for compact binary coalescence events. We also show constraints from $\Delta N_{\rm eff}$ from Planck 2018~\cite{Planck:2018vyg} and projection for CMB-S4~\cite{CMB-S4:2016ple}.}
\label{fig:LISA_LT_sample}
\end{figure}

A representative plot for the frequency dependence of the cosmological and astrophysical signal, along with $\Omega_{\rm GW,n}^\ell$ for $\ell=2$ is given in Fig.~\ref{fig:LISA_LT_sample}.
We choose $C_{\ell=2,\rm{GW}}^c = 10^{-8}$, $\bar{\Omega}^c_{\rm GW}\big\rvert_{\rm peak} = 10^{-8}$, $f_*=4$~mHz (solid olive), and scale the $\bar{\Omega}_{\rm GW}^a = 10^{-9}$ at 10~Hz~\cite{KAGRA:2021kbb} upper limit along with $C_{\ell=2, \rm{GW}}^a = 2\times 10^{-4}$.
For this choice, the noise, rather than the astrophysical source, determines the precision with which $\kappa$ can be measured in LISA or LT.
We also show another benchmark with $f_*=10$~mHz (dot-dashed olive), but otherwise identical parameters.

\begin{figure}
\begin{center}
\includegraphics[width=0.48\textwidth]{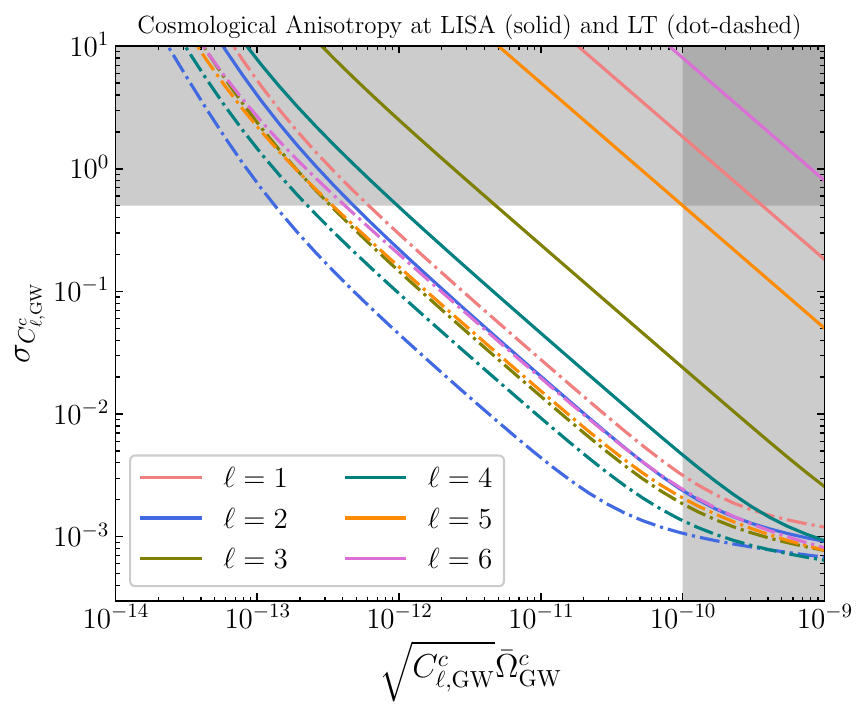}
\includegraphics[width=0.48\textwidth]{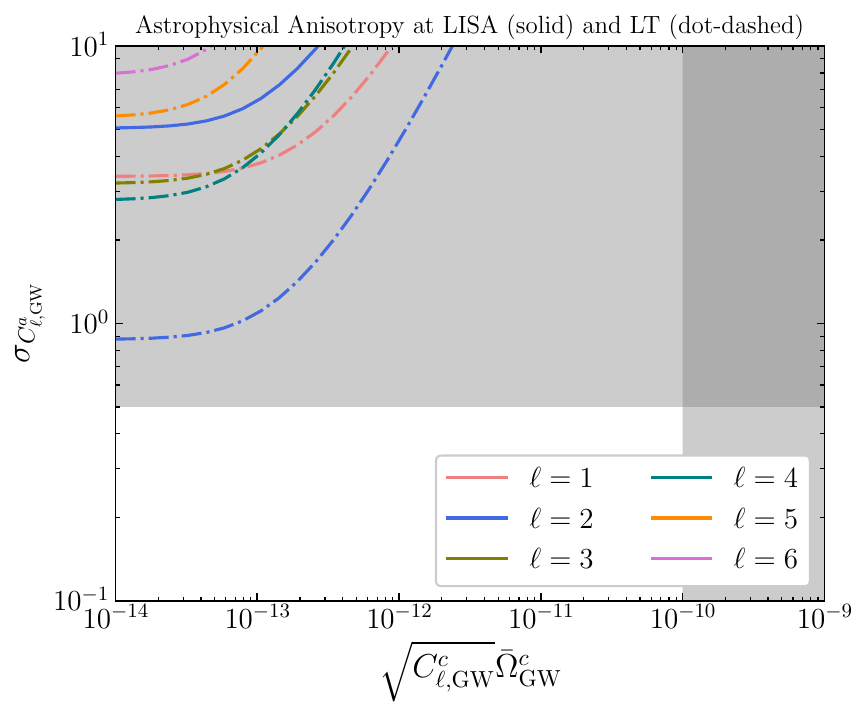}
\end{center}
\caption{Precision on cosmological and astrophysical uncertainties from a Fisher analysis. We assume a PT centered around a frequency of 4~mHz. {\it Left: }Cosmological anisotropy.   {\it Right:} Astrophysical anisotropy.
Via the vertical bands, we show dark radiation isocurvature constraints~\cite{Ghosh:2021axu}, $\sqrt{\omc}>10^{-10}$, since GW behave as dark radiation and can carry non-adiabatic perturbations.
For detectability, we consider only the regions where $\sigma_{\clc}$ and $\sigma_{\cla}$ are smaller than 0.5.
These results illustrate that LISA and Taiji, in combination, can extract anisotropies much better, by accessing more $\ell$ modes, than individually.}
\label{fig:LISA_LT_precision}
\end{figure}

To further quantify the ranges of cosmological anisotropy $\clc$ that are potentially observable, we compute $\sigma_{\clc}$ as a function of various values of $\omc$.
We show the results in Fig.~\ref{fig:LISA_LT_precision} for LISA (solid) and LT (dot-dashed).
We also show the isocurvature constraint~\cite{Ghosh:2021axu} via the vertical band, ruling out the region $\sqrt{\omc}>10^{-10}$, and for detectability, consider only the regions with $\sigma_{\clc},\sigma_{\cla}<0.5$.
Because of the fact that LT has useful sensitivity to more $\ell$-modes, its constraining power is much better than LISA alone.
In particular, Fig.~\ref{fig:LISA_LT_precision} implies that we can observe $\ell=2,3,4$ modes with reasonable precision at LISA, while LT can cover all the modes from $\ell=1$ to $\ell=6$, for cosmological anisotropies.
Since we saturate the LIGO-Virgo upper bound~\cite{KAGRA:2021kbb}, while the actual astrophysical monopole could be lower, the results for $\sigma_{\cla}$ in Fig.~\ref{fig:LISA_LT_precision} suggest that the (extra-galactic) astrophysical anisotropies are likely unobservable at LISA and LT.

Given the discussion in Sec.~\ref{sec:cosmo_contri}, for the case of adiabatic perturbations, our result implies that LISA can observe cosmological SGWB anisotropies ($\ell=2,4$) only for a strong SGWB, such as arising from $(\beta_{\rm PT}/H_{\rm PT})^2\simeq 10$ and $\alpha\simeq 1$.
Nevertheless, weaker PT can still give rise to observable anisotropies if the GW signal carries isocurvature perturbations as discussed toward the end of Sec.~\ref{sec:cosmo_contri}.
LT, on the other hand, can access weaker signals such as with $(\beta_{\rm PT}/H_{\rm PT})^2\simeq 100$ and $\alpha\simeq 1$ along with adiabatic perturbations.

\begin{figure}
\begin{center}
\includegraphics[width=0.48\textwidth]{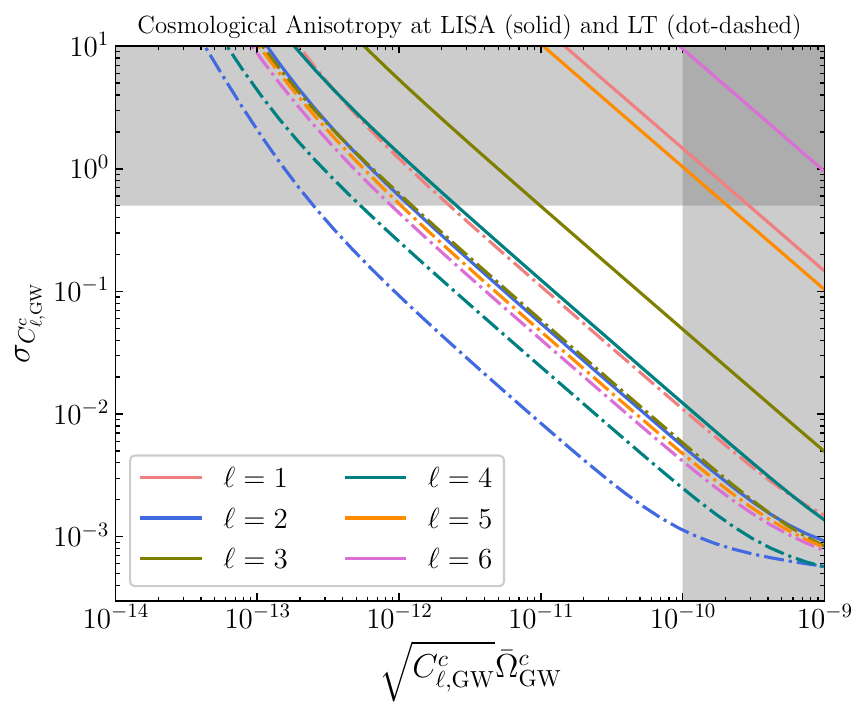}
\includegraphics[width=0.48\textwidth]{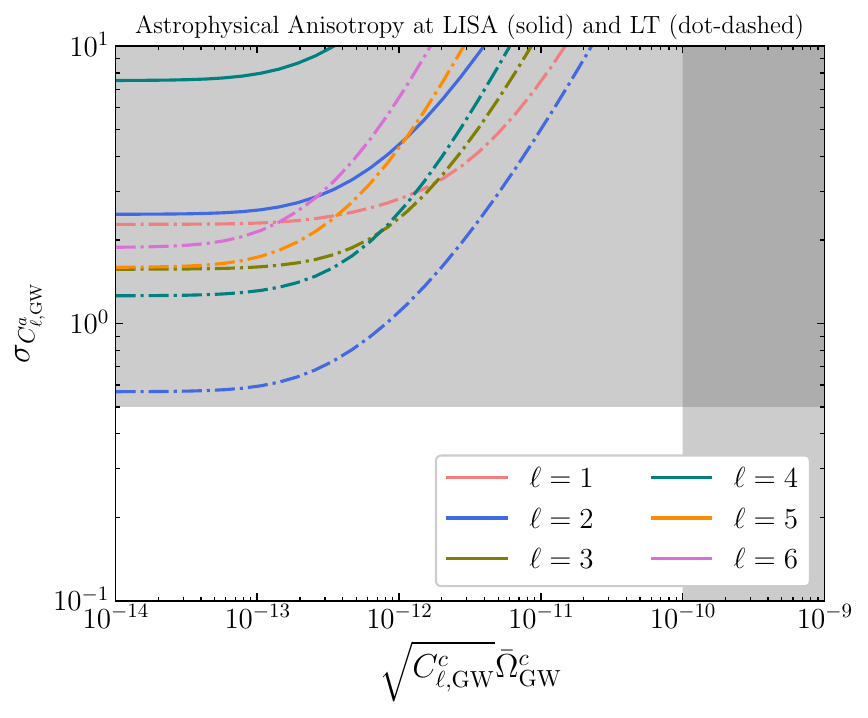}
\end{center}
\caption{Same as Fig.~\ref{fig:LISA_LT_precision}, except the cosmological signal has a peak at 10~mHz, instead of 4~mHz.
As expected based on the dashed olive curve in Fig.~\ref{fig:LISA_LT_sample}, the precision on cosmological anisotropies is worse.
However, LISA and LT are still sensitive to anisotropies in open parts of the parameter space, consistent with bounds on dark radiation isocurvature.}
\label{fig:LISA_LT_precision_10mHz}
\end{figure}

\paragraph{Varying the peak frequency.}
In Fig.~\ref{fig:LISA_LT_precision}, we assumed a PT-generated cosmological signal with a peak frequency of 4~mHz, which is quite close to the frequencies where LISA and LT have the most sensitivity.
To understand how the forecast precision changes as the peak frequency is varied, we now focus on a cosmological signal with a peak frequency at 10~mHz.
The result is shown in Fig.~\ref{fig:LISA_LT_precision_10mHz}.
As expected, the cosmological anisotropies can now be measured with less precision. However, LISA and LT continue to be sensitive to $\ell=2,3,4$ and $\ell=1\dots 6$ modes, respectively.

\subsection{Big Bang Observer}
\begin{figure}
\begin{center}
\includegraphics[width=0.8\textwidth]{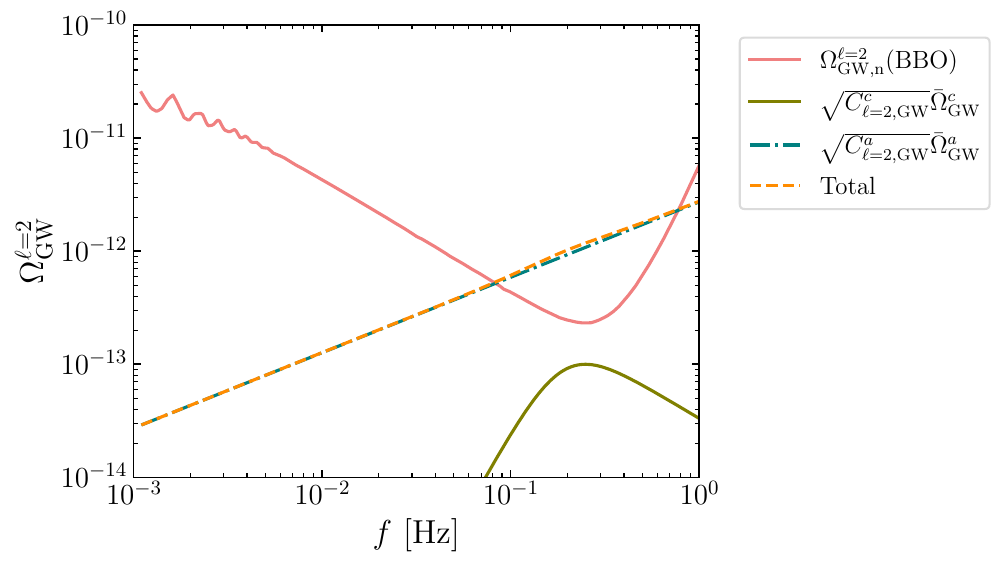}
\end{center}
\caption{
Comparison between astrophysical anisotropy (dot-dashed teal), and cosmological anisotropy (solid olive) with respect to detector sensitivity at BBO (red) for the $\ell=2$ mode. The sum of the two anisotropy sources is shown in dashed orange. We use $C_{\ell=2,\rm{GW}}^c = 10^{-10}$ and $\bar{\Omega}^c_{\rm GW}\big\rvert_{\rm peak} = 10^{-8}$ with $f_* = 0.25$~mHz for the cosmological signal. We also use $C_{\ell=2, \rm{GW}}^a = 2\times 10^{-4}$~\cite{Cusin:2019jhg, Cusin:2019jpv}. As in Fig.~\ref{fig:LISA_LT_sample}, we saturate the astrophysical monopole to its current upper limit, $\bar{\Omega}_{\rm GW}^a = 10^{-9}$ at 10~Hz~\cite{KAGRA:2021kbb}, and then extend it to BBO frequencies using a $f^{2/3}$ scaling.}
\label{fig:BBO_sample}
\end{figure}

To forecast detection prospects at BBO, we use the results of~\cite{Dimastrogiovanni:2022eir} to obtain noise sensitivity curves $\Omega_{\rm GW,n}^\ell$ as a function of frequency.
Since BBO is proposed to have multiple LISA-like triangles, it has powerful sensitivity to both odd and even $\ell$-modes.
A representative plot for the frequency dependence of the cosmological and astrophysical signal, along with $\Omega_{\rm GW,n}^\ell$ for $\ell=2$ is given in Fig.~\ref{fig:BBO_sample}.
We choose $C_{\ell=2,\rm{GW}}^c = 10^{-10}$, $\bar{\Omega}^c_{\rm GW}\big\rvert_{\rm peak} = 10^{-8}$, $f_*=0.25$~Hz, and $\bar{\Omega}_{\rm GW}^a = 9\times 10^{-11}$ at 0.25~Hz, scaling the upper limit in~\cite{KAGRA:2021kbb} with $C_{\ell=2, \rm{GW}}^a = 2\times 10^{-4}$~\cite{Cusin:2019jhg, Cusin:2019jpv}.
As can be seen from Fig.~\ref{fig:BBO_sample}, for this choice the astrophysical component, rather than the detector noise, determines the precision with which cosmological modes can be extracted in BBO.

\begin{figure}
\begin{center}
\includegraphics[width=0.48\textwidth]{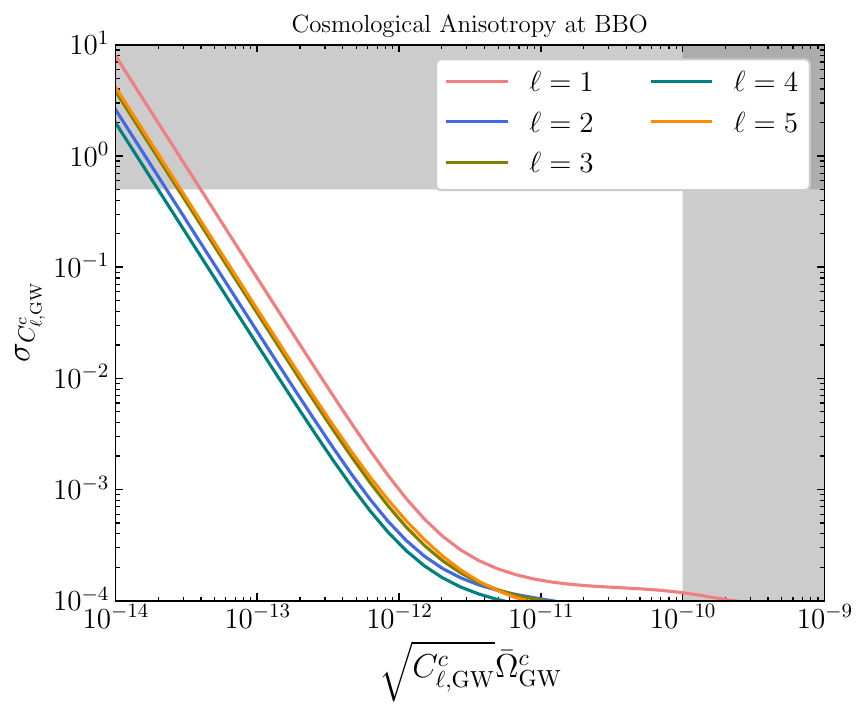}
\includegraphics[width=0.48\textwidth]{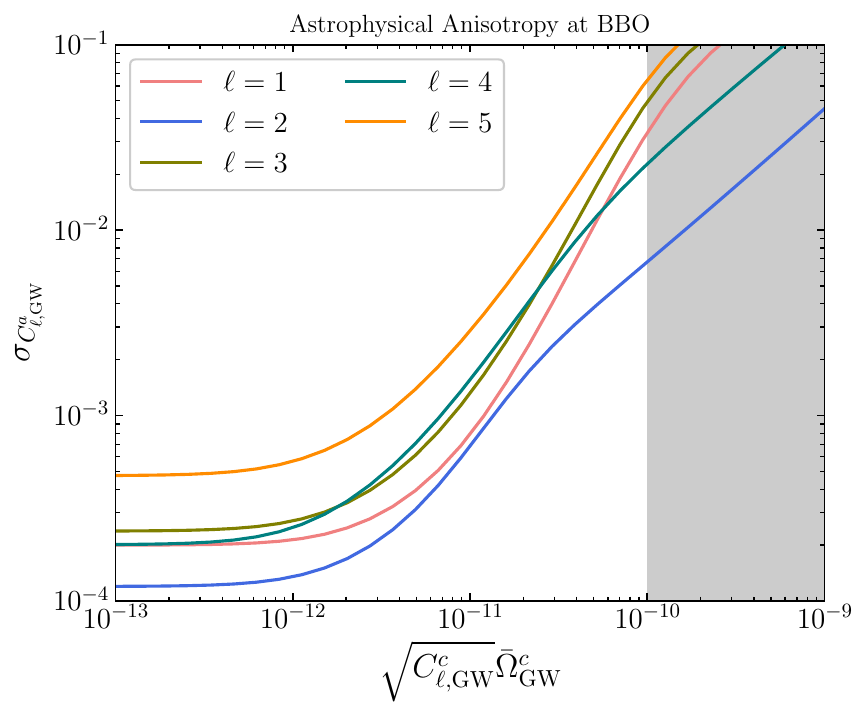}
\end{center}
\caption{Precision on cosmological and astrophysical uncertainties from a Fisher analysis. We assume a PT centered around a frequency of 0.25~Hz. {\it Left: }Cosmological anisotropy.   {\it Right:} Astrophysical anisotropy.
Via the vertical bands, we show dark radiation isocurvature constraints~\cite{Ghosh:2021axu}, $\sqrt{\omc}>10^{-10}$, since GW behaves as dark radiation and can carry non-adiabatic perturbations.
For detectability, we consider only the regions where $\sigma_{\clc}$ and $\sigma_{\cla}$ are smaller than 0.5.
Since BBO is expected to have multiple LISA-like detectors on the same orbit, but at different locations, it can alone be sensitive to both odd and even $\ell$ modes.
In our analysis, we consider modes only up to $\ell=5$, but BBO is expected to powerful sensitivity up to even $\ell\simeq 50$ (see, e.g., Ref.~\cite{Braglia:2021fxn}), although, for these higher $\ell$-modes, spatial shot noise would become important~\cite{Jenkins:2019uzp, Cusin:2019jpv}. 
}
\label{fig:BBO_precision}
\end{figure}

\begin{figure}
\begin{center}
\includegraphics[width=\textwidth]{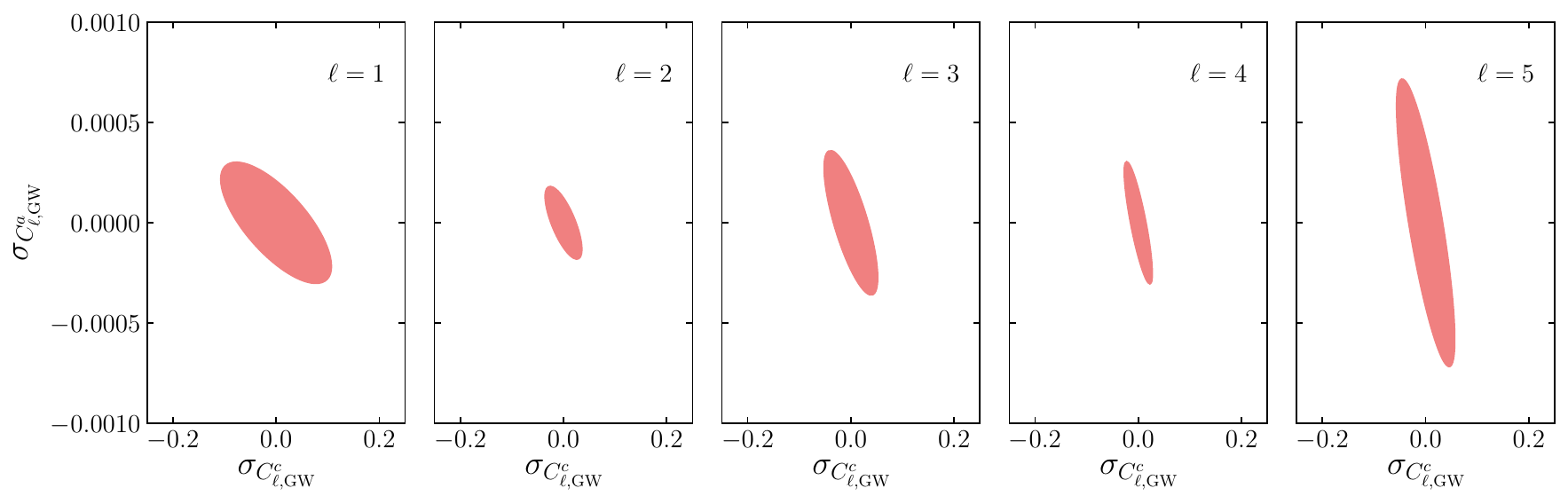}
\end{center}
\caption{Comparison between precision for measuring astrophysical and cosmological anisotropies in the context of BBO for different $\ell$ modes. We choose $\sqrt{\omc}=10^{-13}$. For this choice, the astrophysical anisotropies are much larger than the cosmological ones, as is clear from Fig.~\ref{fig:BBO_sample} for the $\ell=2$ mode.
Nonetheless, we can reliably extract the cosmological anisotropies, albeit with a lower precision compared to the astrophysical ones, i.e., $\sigma_{\cla}\ll \sigma_{\clc}\ll 1$.}
\label{fig:BBO_2D}
\end{figure}

We next compute $\sigma_{\clc}$ as a function of various values of $\omc$.
We show the results in Fig.~\ref{fig:BBO_precision}.
We notice that thanks to the different frequency dependence of the cosmological and the astrophysical signals, both the cosmological and astrophysical signals can be extracted with reasonable precisions.
In particular, in contrast to LISA and LT, BBO can probe astrophysical anisotropies with better than $10\%$ precision.
We also describe the relative precision with which the cosmological and astrophysical signals can be measured in Fig.~\ref{fig:BBO_2D} for $\ell=1$ to $\ell=5$ modes for a benchmark choice $\sqrt{C_{\ell,\rm GW}^c} \bar{\Omega}^c_{\rm GW} = 10^{-13}$.
Since the astrophysical anisotropies are much larger than the cosmological ones for this choice, we get $\sigma_{\cla} \ll \sigma_{\clc}$.
This result also illustrates the fact that BBO has better sensitivity to $\ell=2$ and $\ell=4$ modes compared to the other $\ell$ modes.

Given the discussion in Sec.~\ref{sec:cosmo_contri}, we conclude that BBO can probe adiabatic SGWB anisotropies from a PT corresponding to $(\beta_{\rm PT}/H_{\rm PT})^2 \simeq 10^3$ and $\alpha\simeq 1$ with ${\cal O}(10\%)$ precision.
As we discussed in Sec.~\ref{sec:sources}, properly accounting for correlations between astrophysical and cosmological anisotropies in this case would improve on the expected precision, but we leave this for future work.
For the scenario where SGWB anisotropies carry isocurvature perturbations, BBO can access benchmarks such as $C_{\ell,\rm GW}^c \simeq 10^{-8}$, $(\beta_{\rm PT}/H_{\rm PT})^2 \simeq 10^4$, and $\alpha\simeq 1$ with ${\cal O}(10\%)$ precision as well.

\subsection{Einstein Telescope and Cosmic Explorer}
\begin{figure}
\begin{center}
\includegraphics[width=0.8\textwidth]{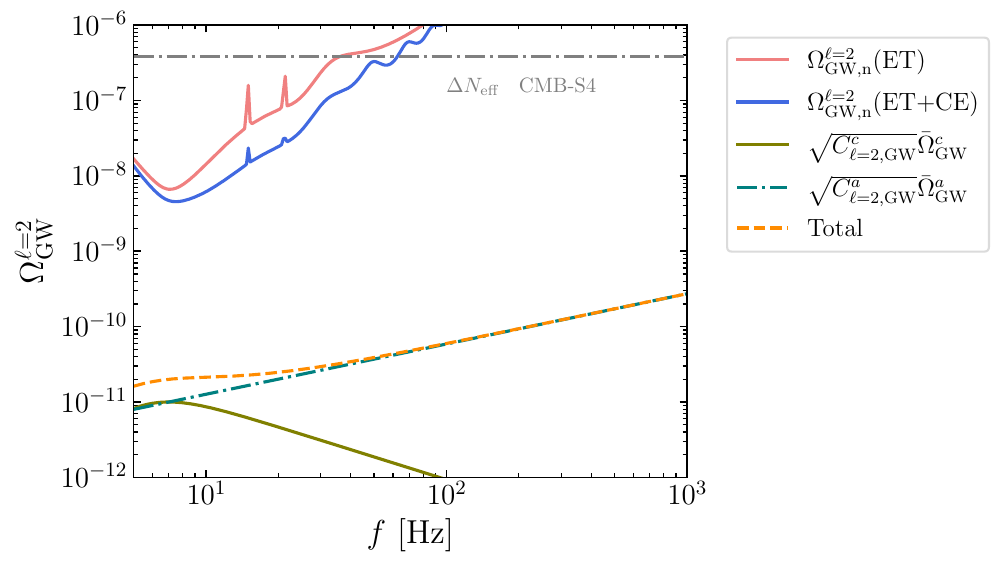}
\end{center}
\caption{
Comparison between astrophysical anisotropy (dot-dashed teal) and cosmological anisotropy (solid olive) with respect to detector sensitivity at ET (red) and ET+CE (blue) for the $\ell=2$ mode. The total anisotropy is shown in dashed orange. We use $C_{\ell=2,\rm{GW}}^c = 10^{-6}$, $\bar{\Omega}^c_{\rm GW}\big\rvert_{\rm peak} = 10^{-8}$ with $f_* = 7$~Hz for the cosmological signal. We also use $C_{\ell=2, \rm{GW}}^a = 2\times 10^{-4}$~\cite{Cusin:2019jhg, Cusin:2019jpv}. As explained in Fig.~\ref{fig:LISA_LT_sample}, we conservatively saturate the astrophysical monopole to its current upper limit~\cite{KAGRA:2021kbb}. We also show projected constraints from $\Delta N_{\rm eff}$ for CMB-S4~\cite{CMB-S4:2016ple}.}
\label{fig:ET_sample}
\end{figure}

We now focus on frequencies in the range of $\sim$1-100~Hz. 
As discussed in Sec.~\ref{sec:sources}, temporal shot noise is significant in these frequencies~\cite{Jenkins:2019uzp}.
While some mitigation strategies have been proposed~\cite{Cusin:2019jpv, Alonso:2020mva}, a dedicated study for Einstein Telescope (ET)/Cosmic Explorer (CE) would be useful in assessing a reduction. 
In this work, we optimistically assume that the temporal shot noise can be fully reduced and consider only the actual astrophysical anisotropies.
With that in mind, to forecast the detection prospects at ET and the combination of ET and CE (ET+CE), we use {\tt schnell} to encode detector specifications and obtain $\Omega_{\rm GW,n}^{\ell}$.
While ET or CE alone does not have powerful sensitivity to odd $\ell$ modes, ET+CE significantly improves that, making the combination sensitive to all modes between $\ell=1$ and $\ell=6$.
A representative plot for the frequency dependence of the cosmological and astrophysical signal, along with $\Omega_{\rm GW,n}^\ell$ for $\ell=2$ is given in Fig.~\ref{fig:ET_sample}.
We choose $C_{\ell=2,\rm{GW}}^c = 10^{-6}$, $\bar{\Omega}^c_{\rm GW}\big\rvert_{\rm peak} = 10^{-8}$, $f_*=7$~Hz, and $\bar{\Omega}_{\rm GW}^a = 10^{-9}$ at 10~Hz~\cite{KAGRA:2021kbb} with $C_{\ell=2, \rm{GW}}^a = 2\times 10^{-4}$~\cite{Cusin:2019jpv, Cusin:2019jhg}.
As can be seen from Fig.~\ref{fig:ET_sample}, for this choice the astrophysical component and the cosmological component are comparable, and we rely on the frequency dependence to separate the two.

\begin{figure}
\begin{center}
\includegraphics[width=0.48\textwidth]{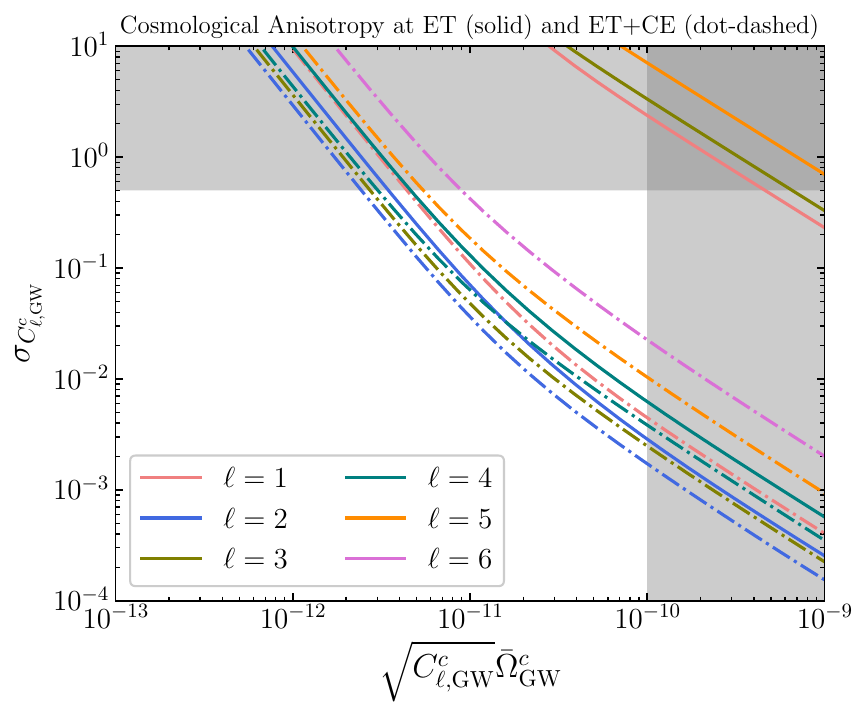}
\includegraphics[width=0.48\textwidth]{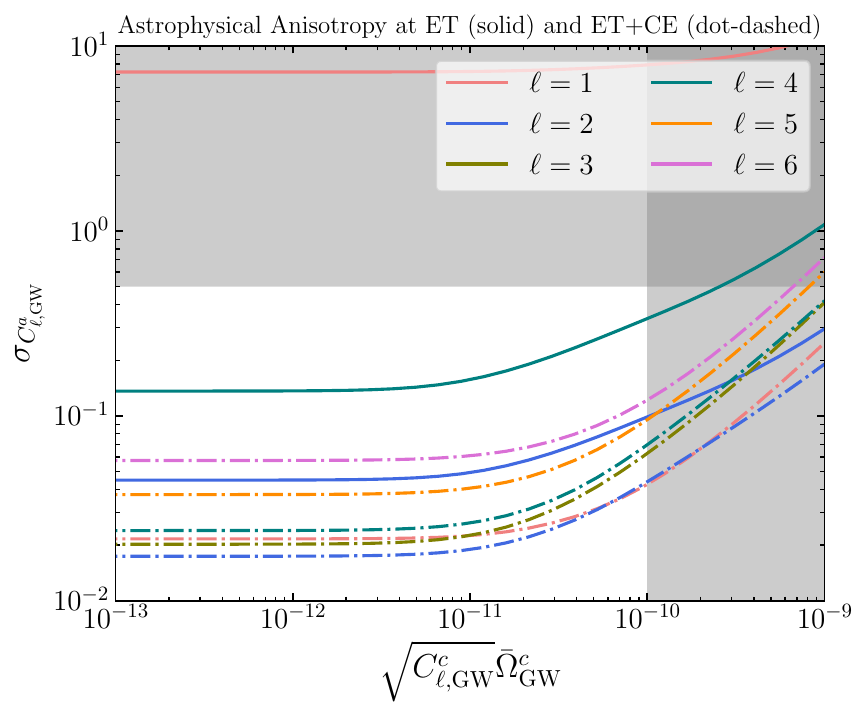}
\end{center}
\caption{Precision on cosmological and astrophysical anisotropies from a Fisher analysis. We assume a PT centered around a frequency of 7~Hz. {\it Left: }Cosmological anisotropy.   {\it Right:} Astrophysical anisotropy.
Via the vertical bands, we show dark radiation isocurvature constraints~\cite{Ghosh:2021axu}, $\sqrt{\omc}>10^{-10}$, since GW behave as dark radiation and can carry non-adiabatic perturbations.
For detectability, we consider only the regions where $\sigma_{\clc}$ and $\sigma_{\cla}$ are smaller than 0.5.
}
\label{fig:ETCE_precision}
\end{figure}

In Fig.~\ref{fig:ETCE_precision}, we show the results for $\sigma_{\clc}$ and $\sigma_{\cla}$ for various choices of cosmological anisotropies $\omc$.
We note that ET alone can provide useful sensitivity to $\ell=2$ and $\ell=4$ modes, while the combination of ET and CE is more powerful, and can extract information from up to $\ell=6$ modes.

In the context of models and focusing on adiabatic perturbations, ET+CE are sensitive to strong SGWB corresponding to $(\beta_{\rm PT}/H_{\rm PT})^2 \simeq 10$ and $\alpha\simeq 1$, and can extract the associated anisotropies with ${\cal O}(10\%)$ precision.
On the other hand, if SGWB carries isocurvature perturbations, ET+CE can access benchmarks such as $C_{\ell,\rm GW}^c \simeq 10^{-6}$, $(\beta_{\rm PT}/H_{\rm PT})^2 \simeq 10^3$, and $\alpha\simeq 1$ with ${\cal O}(10\%)$ precision.

\subsection{Tilt of the Cosmological Anisotropy Power Spectrum}
\begin{figure}
\begin{center}
\includegraphics[width=0.7\textwidth]{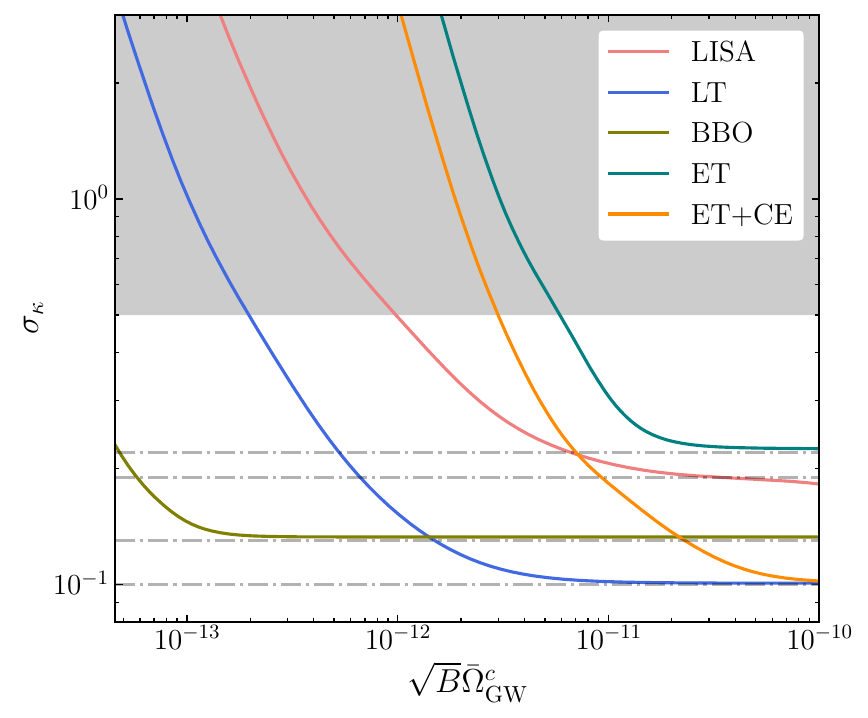}
\end{center}
\caption{Precision on cosmological tilt from a Fisher analysis. 
The dot-dashed gray lines denote the cosmic variance-only precision.
In our analysis, we include $\ell=1\dots 6$ modes, except for BBO, where we include $\ell=1\dots 5$ modes.
Given the detector noise, we find, only $\ell=2,3,4$ modes in LISA; $\ell=1\dots 6$ modes in LT; $\ell=1\dots 5$ modes in BBO; $\ell=2,4$ modes in ET; $\ell=1\dots 6$ modes in ET+CE contribute.
These modes determine the cosmic variance-only precision for each experiment.
For small values of $\sqrt{B}\bar{\Omega}_{\rm GW}^c$, both the cosmic variance term and $\sigma_{\clc}$ in Eq.~\eqref{eq:sig_gamma} contribute. 
However, as $\sqrt{B}\bar{\Omega}_{\rm GW}^c$ gets larger, we can measure the anisotropy corresponding to each $\ell$ mode much better, i.e., $\sigma_{\clc}\ll 1/\sqrt{2\ell+1}$ in that limit. 
For those values, the cosmic variance term in Eq.~\eqref{eq:sig_gamma} dominates and the solid curves asymptote to the dot-dashed lines.}
\label{fig:LISA_LT_tilt}
\end{figure}

Having obtained $\sigma_{\clc}$, we can use Eq.~\eqref{eq:sig_gamma} to compute the precision with which the tilt of the cosmological anisotropy power spectrum can be measured.
To this end, we parametrize the cosmological power spectrum as,
\es{eq:cosmo_power}{
\frac{\ell(\ell+1)}{2\pi}C_{\ell,\rm{GW}}^c \equiv B \ell^\kappa.
}
We then compute $\sigma_\kappa$ as a function of $B$ and $\bar{\Omega}_{\rm GW}^c$.
The result is shown in Fig.~\ref{fig:LISA_LT_tilt}.
Given the flattening of the curves for large $\sqrt{B}\bar{\Omega}_{\rm GW}^c$, we find that at best the spectral tilt can be measured with ${\cal O}(10\%)$ precision by considering modes up to $\ell=6$.
This is because for large enough $\sqrt{B}\bar{\Omega}_{\rm GW}^c$, $\sigma_{\clc}\ll 1$, and it drops out from Eq.~\eqref{eq:sig_gamma}.
In that case, the cosmic variance term dictates the precision.
We show, via dot-dashed horizontal lines, the values of $\sigma_\kappa$ when only cosmic variance is taken into account.
The various curves asymptote to these values, as expected.
We also see how combining two different detectors helps measure $\kappa$ better.
For example, both ET+CE and LT are sensitive to $\ell=1$ through $\ell=6$ modes, and correspondingly the reach in $\sigma_\kappa$ is better than ET alone ($\ell=2, 4$) and LISA alone ($\ell=2,3,4$), respectively.

\section{Conclusion and Discussion}
A discovery of anisotropies imprinted in a cosmologically sourced GW background would be significant, as it can potentially reveal a new map of primordial fluctuations, complementary to what we have learned from CMB anisotropies. 
In particular, a GW background can hide primordial isocurvature fluctuations, a detection of which would give evidence for  multi-field dynamics during the inflationary era.
However, without a dedicated analysis such primordial anisotropies could remain masked by GW anisotropy in the inevitable astrophysical foreground or buried under the detector noise. 
It is thus of paramount importance to ask whether and how well we can unravel the cosmological GW anisotropy from its astrophysical counterpart and the detector noise. 
This is especially relevant given the recent evidence for a stochastic GW background (SGWB) reported by pulsar timing array (PTA) measurements~\cite{NANOGrav:2023gor, Antoniadis:2023ott}.
In recent years, various approaches have been developed in identifying the cosmological signal in the presence of experimental noise and astrophysical foreground, for the isotropic or the `monopole' component of an SGWB. 
In contrast, the analogous work involving anisotropies within an SGWB is still lacking. 

In this work, we took a first step in evaluating the prospect of distinguishing cosmological GW anisotropies from astrophysical sources and detector noises. 
To this end, we make some assumptions regarding astrophysical anisotropies.
The magnitude of the astrophysical foreground is an active area of research.
However, since our purpose here is to understand how well we can extract cosmological anisotropies in the presence of an (extra-galactic) astrophysical foreground, we (conservatively) assume that the monopole of the foreground is the same as the current upper limit on an isotropic SGWB~\cite{KAGRA:2021kbb}.
Of course, the actual astrophysical monopole can be much lower, in which case the prospect of extracting cosmological anisotropies will improve.
For simplicity, we have also assumed that astrophysical and cosmological anisotropies are uncorrelated.
However, if significant correlations are present such as when all sources carry adiabatic fluctuations, this would improve the precision with which the cosmological signal can be extracted, especially at BBO, as discussed in Sec.~\ref{sec:sources}.
The other assumption we make involves shot noise~\cite{Jenkins:2019uzp}.
While the spatial shot noise can be neglected for the $\ell\leq 6$ modes that we consider~\cite{Cusin:2019jpv, Alonso:2020mva}, the temporal (popcorn) shot noise can be significant in the Hz band, relevant for ET/CE.
There are proposals~\cite{Cusin:2019jpv, Alonso:2020mva} for mitigating such shot noise, although dedicated studies for ET/CE would be useful.
Here we optimistically assume that the temporal shot noise can be fully mitigated and only focus on the `true' astrophysical anisotropies.
In this sense, our forecasts for ET/CE are somewhat less robust compared to the forecasts for other experiments.
Once a more detailed mitigation strategy is developed, one can still use a similar Fisher analysis to obtain more robust forecasts.

With these caveats and subtleties in mind, our Fisher forecasts, done on motivated benchmark models for cosmological phase transitions, render a positive outlook.
As examples, we show in Fig.~\ref{fig:LISA_LT_precision} and Fig.~\ref{fig:ETCE_precision} that LISA, LISA-Taiji (LT), and ET+CE combination can measure low-lying ($\ell\leq 6$) modes of an anisotropic cosmological SGWB map with ${\cal O}(10\%)$ error at $1\sigma$ when a first-order phase transition generates a sufficiently strong SGWB.
We find astrophysical anisotropies corresponding to extra-galactic binary neutron star and binary black hole mergers might be unobservable at LISA and LT, but we might be able to detect them in ET+CE.
This is because the astrophysical foreground rises as $f^{2/3}$ as we go to larger frequencies, and therefore the magnitude of the anisotropies, $\delta \Omega_{\rm GW}$, is larger at frequencies relevant for ET+CE.
For BBO frequencies, astrophysical anisotropies can be important, especially as the foreground subtraction may not be perfect. 
We show in Fig.~\ref{fig:BBO_precision} cosmological anisotropies can be constrained with ${\cal O}(10\%)$ precision at $1\sigma$ even when the astrophysical anisotropies are one order of magnitude larger than cosmological ones.
The generic differences in the dependence on frequency between the two sources are key factors aiding the discrimination. 
Detectors with higher angular resolution, in particular a sustained good sensitivity over a range of $\ell$ modes, yield better discriminating power. 
Our forecasts also indicate that the simultaneous operation of multiple detectors probing similar frequencies would be qualitatively more powerful in extracting anisotropies, as evident, for example, by comparing the results of LISA alone with LISA+Taiji.

We also identified the complementarity between monopole and anisotropy discoveries. 
We generally expect that the discovery or hint of the monopole component of a cosmological GW background would precede any evidence for anisotropy, which can provide important information on the signal frequency spectrum.
This can then help improve the sensitivity to the anisotropic component. 
Anticipating this, in our Fisher analysis we have used a fixed frequency dependence for the cosmological signal and obtained forecasts for its anisotropic power.
On the other hand, in the likely scenario that a monopole SGWB gets discovered first, but with debate lingering around whether it is astrophysical or cosmological, further discrimination can be achieved based on their different patterns of anisotropy spectrum. 

To demonstrate the basic idea, we chose the benchmark examples with a cosmological GW source from a first-order phase transition and an astrophysical source from binary BH/NS mergers. 
For cosmological anisotropy, we assumed that it originates from the typical scale-invariant primordial fluctuation, and for its astrophysical counterpart we considered an approximation based on the result in \cite{Cusin:2019jpv}. 
These set the baseline for the shapes of the monopole frequency spectra and the anisotropy spectra in $\ell$. 
While the specifics of the results would differ by considering alternative modeling, our analysis provides a protocol for more general studies, and the positive prospect shown here is expected to hold up for many other cases. 
Given the current uncertainties, in particular, on the astrophysical background, making reasonable assumptions is the best one can do to progress in this direction. 
Practically, one can improve the modeling of astrophysical GW background based on the data LIGO/Virgo/KAGRA is expected to accumulate in the coming years. 
Furthermore, the astrophysical GW anisotropy map is expected to closely trace galaxy distribution and thus can be inferred by correlating with its electromagnetic counterpart observations such as galaxy survey and weak lensing \cite{Baker:2019ync}.

Astrophysical foreground subtraction for cosmological SGWB anisotropy may be improved with various approaches. 
First, as discussed, with the improved sensitivity of the next-generation GW detectors, astrophysical sources can be potentially resolved and subtracted. 
It was recently shown~\cite{Zhou:2022nmt} that such subtraction may be more difficult than previously estimated, leaving a sizable foreground. 
Nevertheless, future dedicated study making use of advancing computing techniques may optimize the effectiveness of such subtraction. 
It may also draw inspiration from techniques developed for analogous study in the context of CMB anisotropy \cite{Tegmark:2003ve, Ichiki:2014qga}. 
However, there are key differences between GWs and CMB, which make discovering a primordial GW anisotropy a unique and new challenge. 
For instance, while there are uncertainties in modeling astrophysical foreground for the CMB, the foreground is known to be most prominent in the galactic plane, and thus galactic masking is effective in greatly reducing the foreground \cite{Ichiki:2014qga}. 
In fact, except for B-modes, galactic masking alone is sufficient for mitigating foregrounds for most CMB anisotropy modes \cite{Ichiki:2014qga, Paoletti:2022anb}. 
On the other hand, the astrophysical foreground for SGWB is generally diffuse, with exceptions such as that originated from galactic white dwarfs.
Furthermore, importantly the CMB signal has well-known features: it dominates the foregrounds in the frequency region around 70 GHz while it fades away at the two sides, and follows a black body distribution in frequency. 
Such known information plays an important role in facilitating foreground subtraction. 
In contrast, for SGWB searches, so far the cosmological signal itself is still a moving target that requires a template hypothesis. 
For GW anisotropy searches, this situation would improve in light of the evidence or discovery of a monopole component. 
The extraction of cosmological CMB anisotropies certainly benefits from the data from observations across multiple frequency bands and good angular resolution. 
In comparison, GW experiments are still at the beginning stages in these regards. 
Future continuous improvements in GW facilities are thus expected to enable better discrimination of a cosmologically sourced anisotropy against its astrophysical counterpart.

With this work as a hopefully inspiring starting point, various future directions can be pursued. For instance, we analyzed a set of GW detectors spanning a wide range of frequencies, from mHz to 10~Hz.
It would be interesting to consider other future GW detectors which may boast low detector noise and high angular resolution such as AEDGE, DECIGO, and TianQin, as well as by the combinations of detectors \cite{Graham:2017lmg, AEDGE:2019nxb, Baker:2019ync, Kawamura:2020pcg, Zhou:2023rop}. 
Furthermore, it would be useful to carry out similar studies in the context of PTAs, especially given the recent evidence for an SGWB~\cite{NANOGrav:2023gor, Antoniadis:2023ott}, but in this case the spatial shot noise would be a major complication~\cite{Sato-Polito:2023spo} which is not present in our current analysis. 
In addition, the synergy of GW experiments targeting different frequency bands is expected to provide more information on the frequency spectral shape of an SGWB signal, thus facilitating the separation between a cosmological and astrophysical source \cite{Lasky:2015lej, Barish:2020vmy}. 
Our study also provides useful guidance on how future experiments may be optimized for identifying a primordial GW anisotropy which may inform the ongoing design/planning.
Finally, we took the relatively simple approach of Fisher analysis to illustrate how well the cosmological signal can be distinguished from astrophysical sources, which may not be poised for optimal signal significance. 
It would also be more optimal to do a combined Fisher analysis that uses frequency and anisotropy spectrum at the same time, instead of analyzing them separately, as we did here. That would be especially useful if somewhat higher $\ell$-modes are considered.
This work, along with related future development, may help fulfill the unique promise of GW anisotropies to shed light on primordial cosmology.

\label{sec:conc}

\section*{Acknowledgment}
We thank Arushi Bodas, Neil Cornish, Alexander Jenkins, and Marc Kamionkowski for helpful discussions.
YC is supported in part by the US Department of Energy (DOE) under award number DE-SC0008541.
SK is supported in part by the U.S. National Science Foundation (NSF) grant PHY-1915314 and the DOE contract DE-AC02-05CH11231. 
RS is supported by NSF grant PHY-2210361 and the Maryland Center for Fundamental Physics.
YT is supported by the NSF grant PHY-2014165.
SK \& YT thank the Mainz Institute of Theoretical Physics of the Cluster of Excellence PRISMA+ (Project ID 39083149) and Aspen Center for Physics (supported by NSF grant PHY-2210452) for their hospitality while this work was in progress.

\bibliographystyle{JHEP}
\bibliography{refs}

\providecommand{\href}[2]{#2}\begingroup\raggedright\begin{thebibliography}{100}

\bibitem{LIGOScientific:2016aoc}
{\bf LIGO Scientific, Virgo} Collaboration, B.~P. Abbott et~al., {\it
  {Observation of Gravitational Waves from a Binary Black Hole Merger}},  {\em
  Phys. Rev. Lett.} {\bf 116} (2016), no.~6 061102,
  [\href{http://arxiv.org/abs/1602.03837}{{\tt arXiv:1602.03837}}].

\bibitem{LIGOScientific:2017ync}
{\bf LIGO Scientific, Virgo, Fermi GBM, INTEGRAL, IceCube, AstroSat Cadmium
  Zinc Telluride Imager Team, IPN, Insight-Hxmt, ANTARES, Swift, AGILE Team,
  1M2H Team, Dark Energy Camera GW-EM, DES, DLT40, GRAWITA, Fermi-LAT, ATCA,
  ASKAP, Las Cumbres Observatory Group, OzGrav, DWF (Deeper Wider Faster
  Program), AST3, CAASTRO, VINROUGE, MASTER, J-GEM, GROWTH, JAGWAR,
  CaltechNRAO, TTU-NRAO, NuSTAR, Pan-STARRS, MAXI Team, TZAC Consortium, KU,
  Nordic Optical Telescope, ePESSTO, GROND, Texas Tech University, SALT Group,
  TOROS, BOOTES, MWA, CALET, IKI-GW Follow-up, H.E.S.S., LOFAR, LWA, HAWC,
  Pierre Auger, ALMA, Euro VLBI Team, Pi of Sky, Chandra Team at McGill
  University, DFN, ATLAS Telescopes, High Time Resolution Universe Survey,
  RIMAS, RATIR, SKA South Africa/MeerKAT} Collaboration, B.~P. Abbott et~al.,
  {\it {Multi-messenger Observations of a Binary Neutron Star Merger}},  {\em
  Astrophys. J. Lett.} {\bf 848} (2017), no.~2 L12,
  [\href{http://arxiv.org/abs/1710.05833}{{\tt arXiv:1710.05833}}].

\bibitem{NANOGrav:2023gor}
{\bf NANOGrav} Collaboration, G.~Agazie et~al., {\it {The NANOGrav 15 yr Data
  Set: Evidence for a Gravitational-wave Background}},  {\em Astrophys. J.
  Lett.} {\bf 951} (2023), no.~1 L8,
  [\href{http://arxiv.org/abs/2306.16213}{{\tt arXiv:2306.16213}}].

\bibitem{Antoniadis:2023ott}
J.~Antoniadis et~al., {\it {The second data release from the European Pulsar
  Timing Array III. Search for gravitational wave signals}},
  \href{http://arxiv.org/abs/2306.16214}{{\tt arXiv:2306.16214}}.

\bibitem{NANOGrav:2023hvm}
{\bf NANOGrav} Collaboration, A.~Afzal et~al., {\it {The NANOGrav 15 yr Data
  Set: Search for Signals from New Physics}},  {\em Astrophys. J. Lett.} {\bf
  951} (2023), no.~1 L11, [\href{http://arxiv.org/abs/2306.16219}{{\tt
  arXiv:2306.16219}}].

\bibitem{Antoniadis:2023xlr}
J.~Antoniadis et~al., {\it {The second data release from the European Pulsar
  Timing Array: V. Implications for massive black holes, dark matter and the
  early Universe}},  \href{http://arxiv.org/abs/2306.16227}{{\tt
  arXiv:2306.16227}}.

\bibitem{Christensen:2018iqi}
N.~Christensen, {\it {Stochastic Gravitational Wave Backgrounds}},  {\em Rept.
  Prog. Phys.} {\bf 82} (2019), no.~1 016903,
  [\href{http://arxiv.org/abs/1811.08797}{{\tt arXiv:1811.08797}}].

\bibitem{Renzini:2022alw}
A.~I. Renzini, B.~Goncharov, A.~C. Jenkins, and P.~M. Meyers, {\it {Stochastic
  Gravitational-Wave Backgrounds: Current Detection Efforts and Future
  Prospects}},  {\em Galaxies} {\bf 10} (2022), no.~1 34,
  [\href{http://arxiv.org/abs/2202.00178}{{\tt arXiv:2202.00178}}].

\bibitem{vanRemortel:2022fkb}
N.~van Remortel, K.~Janssens, and K.~Turbang, {\it {Stochastic gravitational
  wave background: Methods and implications}},  {\em Prog. Part. Nucl. Phys.}
  {\bf 128} (2023) 104003, [\href{http://arxiv.org/abs/2210.00761}{{\tt
  arXiv:2210.00761}}].

\bibitem{LIGOAplus}
{\it {The $A^+$ design curve}},  tech. rep., {2018}.
\newblock \url{https://dcc.ligo.org/LIGO-T1800042/public}.

\bibitem{Boyle:2005se}
L.~A. Boyle and P.~J. Steinhardt, {\it {Probing the early universe with
  inflationary gravitational waves}},  {\em Phys. Rev. D} {\bf 77} (2008)
  063504, [\href{http://arxiv.org/abs/astro-ph/0512014}{{\tt
  astro-ph/0512014}}].

\bibitem{Boyle:2007zx}
L.~A. Boyle and A.~Buonanno, {\it {Relating gravitational wave constraints from
  primordial nucleosynthesis, pulsar timing, laser interferometers, and the
  CMB: Implications for the early Universe}},  {\em Phys. Rev. D} {\bf 78}
  (2008) 043531, [\href{http://arxiv.org/abs/0708.2279}{{\tt
  arXiv:0708.2279}}].

\bibitem{Caprini:2018mtu}
C.~Caprini and D.~G. Figueroa, {\it {Cosmological Backgrounds of Gravitational
  Waves}},  {\em Class. Quant. Grav.} {\bf 35} (2018), no.~16 163001,
  [\href{http://arxiv.org/abs/1801.04268}{{\tt arXiv:1801.04268}}].

\bibitem{Caldwell:2022qsj}
R.~Caldwell et~al., {\it {Detection of Early-Universe Gravitational Wave
  Signatures and Fundamental Physics}},
  \href{http://arxiv.org/abs/2203.07972}{{\tt arXiv:2203.07972}}.

\bibitem{Geller:2018mwu}
M.~Geller, A.~Hook, R.~Sundrum, and Y.~Tsai, {\it {Primordial Anisotropies in
  the Gravitational Wave Background from Cosmological Phase Transitions}},
  {\em Phys. Rev. Lett.} {\bf 121} (2018), no.~20 201303,
  [\href{http://arxiv.org/abs/1803.10780}{{\tt arXiv:1803.10780}}].

\bibitem{Bethke:2013aba}
L.~Bethke, D.~G. Figueroa, and A.~Rajantie, {\it {Anisotropies in the
  Gravitational Wave Background from Preheating}},  {\em Phys. Rev. Lett.} {\bf
  111} (2013), no.~1 011301, [\href{http://arxiv.org/abs/1304.2657}{{\tt
  arXiv:1304.2657}}].

\bibitem{Bethke:2013vca}
L.~Bethke, D.~G. Figueroa, and A.~Rajantie, {\it {On the Anisotropy of the
  Gravitational Wave Background from Massless Preheating}},  {\em JCAP} {\bf
  06} (2014) 047, [\href{http://arxiv.org/abs/1309.1148}{{\tt
  arXiv:1309.1148}}].

\bibitem{Dimastrogiovanni:2019bfl}
E.~Dimastrogiovanni, M.~Fasiello, and G.~Tasinato, {\it {Searching for Fossil
  Fields in the Gravity Sector}},  {\em Phys. Rev. Lett.} {\bf 124} (2020),
  no.~6 061302, [\href{http://arxiv.org/abs/1906.07204}{{\tt
  arXiv:1906.07204}}].

\bibitem{Adshead:2020bji}
P.~Adshead, N.~Afshordi, E.~Dimastrogiovanni, M.~Fasiello, E.~A. Lim, and
  G.~Tasinato, {\it {Multimessenger cosmology: Correlating cosmic microwave
  background and stochastic gravitational wave background measurements}},  {\em
  Phys. Rev. D} {\bf 103} (2021), no.~2 023532,
  [\href{http://arxiv.org/abs/2004.06619}{{\tt arXiv:2004.06619}}].

\bibitem{Dimastrogiovanni:2021mfs}
E.~Dimastrogiovanni, M.~Fasiello, A.~Malhotra, P.~D. Meerburg, and G.~Orlando,
  {\it {Testing the early universe with anisotropies of the gravitational wave
  background}},  {\em JCAP} {\bf 02} (2022), no.~02 040,
  [\href{http://arxiv.org/abs/2109.03077}{{\tt arXiv:2109.03077}}].

\bibitem{Kumar:2021ffi}
S.~Kumar, R.~Sundrum, and Y.~Tsai, {\it {Non-Gaussian stochastic gravitational
  waves from phase transitions}},  {\em JHEP} {\bf 11} (2021) 107,
  [\href{http://arxiv.org/abs/2102.05665}{{\tt arXiv:2102.05665}}].

\bibitem{Bodas:2022zca}
A.~Bodas and R.~Sundrum, {\it {Primordial clocks within stochastic
  gravitational wave anisotropies}},  {\em JCAP} {\bf 10} (2022) 012,
  [\href{http://arxiv.org/abs/2205.04482}{{\tt arXiv:2205.04482}}].

\bibitem{Bodas:2022urf}
A.~Bodas and R.~Sundrum, {\it {Large Primordial Fluctuations in Gravitational
  Waves from Phase Transitions}},  \href{http://arxiv.org/abs/2211.09301}{{\tt
  arXiv:2211.09301}}.

\bibitem{Contaldi:2016koz}
C.~R. Contaldi, {\it {Anisotropies of Gravitational Wave Backgrounds: A Line Of
  Sight Approach}},  {\em Phys. Lett. B} {\bf 771} (2017) 9--12,
  [\href{http://arxiv.org/abs/1609.08168}{{\tt arXiv:1609.08168}}].

\bibitem{Bartolo:2019yeu}
N.~Bartolo, D.~Bertacca, S.~Matarrese, M.~Peloso, A.~Ricciardone, A.~Riotto,
  and G.~Tasinato, {\it {Characterizing the cosmological gravitational wave
  background: Anisotropies and non-Gaussianity}},  {\em Phys. Rev. D} {\bf 102}
  (2020), no.~2 023527, [\href{http://arxiv.org/abs/1912.09433}{{\tt
  arXiv:1912.09433}}].

\bibitem{Bartolo:2019oiq}
N.~Bartolo, D.~Bertacca, S.~Matarrese, M.~Peloso, A.~Ricciardone, A.~Riotto,
  and G.~Tasinato, {\it {Anisotropies and non-Gaussianity of the Cosmological
  Gravitational Wave Background}},  {\em Phys. Rev. D} {\bf 100} (2019), no.~12
  121501, [\href{http://arxiv.org/abs/1908.00527}{{\tt arXiv:1908.00527}}].

\bibitem{Olmez:2011cg}
S.~Olmez, V.~Mandic, and X.~Siemens, {\it {Anisotropies in the
  Gravitational-Wave Stochastic Background}},  {\em JCAP} {\bf 07} (2012) 009,
  [\href{http://arxiv.org/abs/1106.5555}{{\tt arXiv:1106.5555}}].

\bibitem{Kuroyanagi:2016ugi}
S.~Kuroyanagi, K.~Takahashi, N.~Yonemaru, and H.~Kumamoto, {\it {Anisotropies
  in the gravitational wave background as a probe of the cosmic string
  network}},  {\em Phys. Rev. D} {\bf 95} (2017), no.~4 043531,
  [\href{http://arxiv.org/abs/1604.00332}{{\tt arXiv:1604.00332}}].

\bibitem{Jenkins:2018nty}
A.~C. Jenkins and M.~Sakellariadou, {\it {Anisotropies in the stochastic
  gravitational-wave background: Formalism and the cosmic string case}},  {\em
  Phys. Rev. D} {\bf 98} (2018), no.~6 063509,
  [\href{http://arxiv.org/abs/1802.06046}{{\tt arXiv:1802.06046}}].

\bibitem{Cai:2021dgx}
R.-G. Cai, Z.-K. Guo, and J.~Liu, {\it {A New Picture of Cosmic String
  Evolution and Anisotropic Stochastic Gravitational-Wave Background}},
  \href{http://arxiv.org/abs/2112.10131}{{\tt arXiv:2112.10131}}.

\bibitem{Parida:2015fma}
A.~Parida, S.~Mitra, and S.~Jhingan, {\it {Component Separation of a Isotropic
  Gravitational Wave Background}},  {\em JCAP} {\bf 04} (2016) 024,
  [\href{http://arxiv.org/abs/1510.07994}{{\tt arXiv:1510.07994}}].

\bibitem{Karnesis:2019mph}
N.~Karnesis, M.~Lilley, and A.~Petiteau, {\it {Assessing the detectability of a
  Stochastic Gravitational Wave Background with LISA, using an excess of power
  approach}},  {\em Class. Quant. Grav.} {\bf 37} (2020), no.~21 215017,
  [\href{http://arxiv.org/abs/1906.09027}{{\tt arXiv:1906.09027}}].

\bibitem{Caprini:2019pxz}
C.~Caprini, D.~G. Figueroa, R.~Flauger, G.~Nardini, M.~Peloso, M.~Pieroni,
  A.~Ricciardone, and G.~Tasinato, {\it {Reconstructing the spectral shape of a
  stochastic gravitational wave background with LISA}},  {\em JCAP} {\bf 11}
  (2019) 017, [\href{http://arxiv.org/abs/1906.09244}{{\tt arXiv:1906.09244}}].

\bibitem{Biscoveanu:2020gds}
S.~Biscoveanu, C.~Talbot, E.~Thrane, and R.~Smith, {\it {Measuring the
  primordial gravitational-wave background in the presence of astrophysical
  foregrounds}},  {\em Phys. Rev. Lett.} {\bf 125} (2020) 241101,
  [\href{http://arxiv.org/abs/2009.04418}{{\tt arXiv:2009.04418}}].

\bibitem{Flauger:2020qyi}
R.~Flauger, N.~Karnesis, G.~Nardini, M.~Pieroni, A.~Ricciardone, and
  J.~Torrado, {\it {Improved reconstruction of a stochastic gravitational wave
  background with LISA}},  {\em JCAP} {\bf 01} (2021) 059,
  [\href{http://arxiv.org/abs/2009.11845}{{\tt arXiv:2009.11845}}].

\bibitem{Pieroni:2020rob}
M.~Pieroni and E.~Barausse, {\it {Foreground cleaning and template-free
  stochastic background extraction for LISA}},  {\em JCAP} {\bf 07} (2020) 021,
  [\href{http://arxiv.org/abs/2004.01135}{{\tt arXiv:2004.01135}}]. [Erratum:
  JCAP 09, E01 (2020)].

\bibitem{Barish:2020vmy}
B.~C. Barish, S.~Bird, and Y.~Cui, {\it {Impact of a midband gravitational wave
  experiment on detectability of cosmological stochastic gravitational wave
  backgrounds}},  {\em Phys. Rev. D} {\bf 103} (2021), no.~12 123541,
  [\href{http://arxiv.org/abs/2012.07874}{{\tt arXiv:2012.07874}}].

\bibitem{Martinovic:2020hru}
K.~Martinovic, P.~M. Meyers, M.~Sakellariadou, and N.~Christensen, {\it
  {Simultaneous estimation of astrophysical and cosmological stochastic
  gravitational-wave backgrounds with terrestrial detectors}},  {\em Phys. Rev.
  D} {\bf 103} (2021), no.~4 043023,
  [\href{http://arxiv.org/abs/2011.05697}{{\tt arXiv:2011.05697}}].

\bibitem{Boileau:2021sni}
G.~Boileau, A.~Lamberts, N.~J. Cornish, and R.~Meyer, {\it {Spectral separation
  of the stochastic gravitational-wave background for LISA in the context of a
  modulated Galactic foreground}},  {\em Mon. Not. Roy. Astron. Soc.} {\bf 508}
  (2021), no.~1 803--826, [\href{http://arxiv.org/abs/2105.04283}{{\tt
  arXiv:2105.04283}}]. [Erratum: Mon.Not.Roy.Astron.Soc. 508, 5554--5555
  (2021)].

\bibitem{Boileau:2021gbr}
G.~Boileau, A.~C. Jenkins, M.~Sakellariadou, R.~Meyer, and N.~Christensen, {\it
  {Ability of LISA to detect a gravitational-wave background of cosmological
  origin: The cosmic string case}},  {\em Phys. Rev. D} {\bf 105} (2022), no.~2
  023510, [\href{http://arxiv.org/abs/2109.06552}{{\tt arXiv:2109.06552}}].

\bibitem{Gowling:2021gcy}
C.~Gowling and M.~Hindmarsh, {\it {Observational prospects for phase
  transitions at LISA: Fisher matrix analysis}},  {\em JCAP} {\bf 10} (2021)
  039, [\href{http://arxiv.org/abs/2106.05984}{{\tt arXiv:2106.05984}}].

\bibitem{Lewicki:2021kmu}
M.~Lewicki and V.~Vaskonen, {\it {Impact of LIGO-Virgo binaries on
  gravitational wave background searches}},
  \href{http://arxiv.org/abs/2111.05847}{{\tt arXiv:2111.05847}}.

\bibitem{Boileau:2022ter}
G.~Boileau, N.~Christensen, C.~Gowling, M.~Hindmarsh, and R.~Meyer, {\it
  {Prospects for LISA to detect a gravitational-wave background from first
  order phase transitions}},  \href{http://arxiv.org/abs/2209.13277}{{\tt
  arXiv:2209.13277}}.

\bibitem{Boileau:2022uos}
G.~Boileau, N.~Christensen, and R.~Meyer, {\it {Figures of merit for a
  stochastic gravitational-wave background measurement by LISA: Implications of
  LISA Pathfinder noise correlations}},  {\em Phys. Rev. D} {\bf 106} (2022),
  no.~6 063025, [\href{http://arxiv.org/abs/2204.03867}{{\tt
  arXiv:2204.03867}}].

\bibitem{Racco:2022bwj}
D.~Racco and D.~Poletti, {\it {Precision cosmology with primordial GW
  backgrounds in presence of astrophysical foregrounds}},  {\em JCAP} {\bf 04}
  (2023) 054, [\href{http://arxiv.org/abs/2212.06602}{{\tt arXiv:2212.06602}}].

\bibitem{Baghi:2023qnq}
Q.~Baghi, N.~Karnesis, J.-B. Bayle, M.~Besan\c{c}on, and H.~Inchausp\'e, {\it
  {Uncovering gravitational-wave backgrounds from noises of unknown shape with
  LISA}},  {\em JCAP} {\bf 04} (2023) 066,
  [\href{http://arxiv.org/abs/2302.12573}{{\tt arXiv:2302.12573}}].

\bibitem{Mukherjee:2019oma}
S.~Mukherjee and J.~Silk, {\it {Time-dependence of the astrophysical stochastic
  gravitational wave background}},  {\em Mon. Not. Roy. Astron. Soc.} {\bf 491}
  (2020), no.~4 4690--4701, [\href{http://arxiv.org/abs/1912.07657}{{\tt
  arXiv:1912.07657}}].

\bibitem{Mukherjee:2020jxa}
S.~Mukherjee and J.~Silk, {\it {Fundamental physics using the temporal
  gravitational wave background}},  {\em Phys. Rev. D} {\bf 104} (2021), no.~6
  063518, [\href{http://arxiv.org/abs/2008.01082}{{\tt arXiv:2008.01082}}].

\bibitem{Caprini:2009fx}
C.~Caprini, R.~Durrer, T.~Konstandin, and G.~Servant, {\it {General Properties
  of the Gravitational Wave Spectrum from Phase Transitions}},  {\em Phys. Rev.
  D} {\bf 79} (2009) 083519, [\href{http://arxiv.org/abs/0901.1661}{{\tt
  arXiv:0901.1661}}].

\bibitem{Hindmarsh:2017gnf}
M.~Hindmarsh, S.~J. Huber, K.~Rummukainen, and D.~J. Weir, {\it {Shape of the
  acoustic gravitational wave power spectrum from a first order phase
  transition}},  {\em Phys. Rev. D} {\bf 96} (2017), no.~10 103520,
  [\href{http://arxiv.org/abs/1704.05871}{{\tt arXiv:1704.05871}}]. [Erratum:
  Phys.Rev.D 101, 089902 (2020)].

\bibitem{Caprini:2019egz}
C.~Caprini et~al., {\it {Detecting gravitational waves from cosmological phase
  transitions with LISA: an update}},  {\em JCAP} {\bf 03} (2020) 024,
  [\href{http://arxiv.org/abs/1910.13125}{{\tt arXiv:1910.13125}}].

\bibitem{Ellis:2018mja}
J.~Ellis, M.~Lewicki, and J.~M. No, {\it {On the Maximal Strength of a
  First-Order Electroweak Phase Transition and its Gravitational Wave Signal}},
   {\em JCAP} {\bf 04} (2019) 003, [\href{http://arxiv.org/abs/1809.08242}{{\tt
  arXiv:1809.08242}}].

\bibitem{Ellis:2019oqb}
J.~Ellis, M.~Lewicki, J.~M. No, and V.~Vaskonen, {\it {Gravitational wave
  energy budget in strongly supercooled phase transitions}},  {\em JCAP} {\bf
  06} (2019) 024, [\href{http://arxiv.org/abs/1903.09642}{{\tt
  arXiv:1903.09642}}].

\bibitem{Phinney:2001di}
E.~S. Phinney, {\it {A Practical theorem on gravitational wave backgrounds}},
  \href{http://arxiv.org/abs/astro-ph/0108028}{{\tt astro-ph/0108028}}.

\bibitem{Farmer:2003pa}
A.~J. Farmer and E.~S. Phinney, {\it {The gravitational wave background from
  cosmological compact binaries}},  {\em Mon. Not. Roy. Astron. Soc.} {\bf 346}
  (2003) 1197, [\href{http://arxiv.org/abs/astro-ph/0304393}{{\tt
  astro-ph/0304393}}].

\bibitem{Cusin:2017fwz}
G.~Cusin, C.~Pitrou, and J.-P. Uzan, {\it {Anisotropy of the astrophysical
  gravitational wave background: Analytic expression of the angular power
  spectrum and correlation with cosmological observations}},  {\em Phys. Rev.
  D} {\bf 96} (2017), no.~10 103019,
  [\href{http://arxiv.org/abs/1704.06184}{{\tt arXiv:1704.06184}}].

\bibitem{Cusin:2018rsq}
G.~Cusin, I.~Dvorkin, C.~Pitrou, and J.-P. Uzan, {\it {First predictions of the
  angular power spectrum of the astrophysical gravitational wave background}},
  {\em Phys. Rev. Lett.} {\bf 120} (2018) 231101,
  [\href{http://arxiv.org/abs/1803.03236}{{\tt arXiv:1803.03236}}].

\bibitem{Jenkins:2018kxc}
A.~C. Jenkins, R.~O'Shaughnessy, M.~Sakellariadou, and D.~Wysocki, {\it
  {Anisotropies in the astrophysical gravitational-wave background: The impact
  of black hole distributions}},  {\em Phys. Rev. Lett.} {\bf 122} (2019),
  no.~11 111101, [\href{http://arxiv.org/abs/1810.13435}{{\tt
  arXiv:1810.13435}}].

\bibitem{Cusin:2019jhg}
G.~Cusin, I.~Dvorkin, C.~Pitrou, and J.-P. Uzan, {\it {Stochastic gravitational
  wave background anisotropies in the mHz band: astrophysical dependencies}},
  {\em Mon. Not. Roy. Astron. Soc.} {\bf 493} (2020), no.~1 L1--L5,
  [\href{http://arxiv.org/abs/1904.07757}{{\tt arXiv:1904.07757}}].

\bibitem{Cusin:2019jpv}
G.~Cusin, I.~Dvorkin, C.~Pitrou, and J.-P. Uzan, {\it {Properties of the
  stochastic astrophysical gravitational wave background: astrophysical sources
  dependencies}},  {\em Phys. Rev. D} {\bf 100} (2019), no.~6 063004,
  [\href{http://arxiv.org/abs/1904.07797}{{\tt arXiv:1904.07797}}].

\bibitem{Bellomo:2021mer}
N.~Bellomo, D.~Bertacca, A.~C. Jenkins, S.~Matarrese, A.~Raccanelli,
  T.~Regimbau, A.~Ricciardone, and M.~Sakellariadou, {\it {CLASS\_GWB: robust
  modeling of the astrophysical gravitational wave background anisotropies}},
  {\em JCAP} {\bf 06} (2022), no.~06 030,
  [\href{http://arxiv.org/abs/2110.15059}{{\tt arXiv:2110.15059}}].

\bibitem{Chen:2010xka}
X.~Chen, {\it {Primordial Non-Gaussianities from Inflation Models}},  {\em Adv.
  Astron.} {\bf 2010} (2010) 638979,
  [\href{http://arxiv.org/abs/1002.1416}{{\tt arXiv:1002.1416}}].

\bibitem{Chluba:2015bqa}
J.~Chluba, J.~Hamann, and S.~P. Patil, {\it {Features and New Physical Scales
  in Primordial Observables: Theory and Observation}},  {\em Int. J. Mod. Phys.
  D} {\bf 24} (2015), no.~10 1530023,
  [\href{http://arxiv.org/abs/1505.01834}{{\tt arXiv:1505.01834}}].

\bibitem{KAGRA:2021kbb}
{\bf KAGRA, Virgo, LIGO Scientific} Collaboration, R.~Abbott et~al., {\it
  {Upper limits on the isotropic gravitational-wave background from Advanced
  LIGO and Advanced Virgo\textquoteright{}s third observing run}},  {\em Phys.
  Rev. D} {\bf 104} (2021), no.~2 022004,
  [\href{http://arxiv.org/abs/2101.12130}{{\tt arXiv:2101.12130}}].

\bibitem{Planck:2018jri}
{\bf Planck} Collaboration, Y.~Akrami et~al., {\it {Planck 2018 results. X.
  Constraints on inflation}},  {\em Astron. Astrophys.} {\bf 641} (2020) A10,
  [\href{http://arxiv.org/abs/1807.06211}{{\tt arXiv:1807.06211}}].

\bibitem{Ghosh:2021axu}
S.~Ghosh, S.~Kumar, and Y.~Tsai, {\it {Free-streaming and coupled dark
  radiation isocurvature perturbations: constraints and application to the
  Hubble tension}},  {\em JCAP} {\bf 05} (2022), no.~05 014,
  [\href{http://arxiv.org/abs/2107.09076}{{\tt arXiv:2107.09076}}].

\bibitem{Regimbau:2016ike}
T.~Regimbau, M.~Evans, N.~Christensen, E.~Katsavounidis, B.~Sathyaprakash, and
  S.~Vitale, {\it {Digging deeper: Observing primordial gravitational waves
  below the binary black hole produced stochastic background}},  {\em Phys.
  Rev. Lett.} {\bf 118} (2017), no.~15 151105,
  [\href{http://arxiv.org/abs/1611.08943}{{\tt arXiv:1611.08943}}].

\bibitem{Sachdev:2020bkk}
S.~Sachdev, T.~Regimbau, and B.~S. Sathyaprakash, {\it {Subtracting compact
  binary foreground sources to reveal primordial gravitational-wave
  backgrounds}},  {\em Phys. Rev. D} {\bf 102} (2020), no.~2 024051,
  [\href{http://arxiv.org/abs/2002.05365}{{\tt arXiv:2002.05365}}].

\bibitem{Zhou:2022nmt}
B.~Zhou, L.~Reali, E.~Berti, M.~\c{C}al\i{}\c{s}kan, C.~Creque-Sarbinowski,
  M.~Kamionkowski, and B.~S. Sathyaprakash, {\it {Subtracting Compact Binary
  Foregrounds to Search for Subdominant Gravitational-Wave Backgrounds in
  Next-Generation Ground-Based Observatories}},
  \href{http://arxiv.org/abs/2209.01310}{{\tt arXiv:2209.01310}}.

\bibitem{Zhou:2022otw}
B.~Zhou, L.~Reali, E.~Berti, M.~\c{C}al\i{}\c{s}kan, C.~Creque-Sarbinowski,
  M.~Kamionkowski, and B.~S. Sathyaprakash, {\it {Compact Binary Foreground
  Subtraction in Next-Generation Ground-Based Observatories}},
  \href{http://arxiv.org/abs/2209.01221}{{\tt arXiv:2209.01221}}.

\bibitem{Cutler:2005qq}
C.~Cutler and J.~Harms, {\it {BBO and the neutron-star-binary subtraction
  problem}},  {\em Phys. Rev. D} {\bf 73} (2006) 042001,
  [\href{http://arxiv.org/abs/gr-qc/0511092}{{\tt gr-qc/0511092}}].

\bibitem{Braglia:2021fxn}
M.~Braglia and S.~Kuroyanagi, {\it {Probing prerecombination physics by the
  cross-correlation of stochastic gravitational waves and CMB anisotropies}},
  {\em Phys. Rev. D} {\bf 104} (2021), no.~12 123547,
  [\href{http://arxiv.org/abs/2106.03786}{{\tt arXiv:2106.03786}}].

\bibitem{LISACosmologyWorkingGroup:2022kbp}
{\bf LISA Cosmology Working Group} Collaboration, N.~Bartolo et~al., {\it
  {Probing Anisotropies of the Stochastic Gravitational Wave Background with
  LISA}},  \href{http://arxiv.org/abs/2201.08782}{{\tt arXiv:2201.08782}}.

\bibitem{Ali-Haimoud:2020iyz}
Y.~Ali-Ha\"\i{}moud, T.~L. Smith, and C.~M.~F. Mingarelli, {\it {Insights into
  searches for anisotropies in the nanohertz gravitational-wave background}},
  {\em Phys. Rev. D} {\bf 103} (2021), no.~4 042009,
  [\href{http://arxiv.org/abs/2010.13958}{{\tt arXiv:2010.13958}}].

\bibitem{KAGRA:2021mth}
{\bf KAGRA, Virgo, LIGO Scientific} Collaboration, R.~Abbott et~al., {\it
  {Search for anisotropic gravitational-wave backgrounds using data from
  Advanced LIGO and Advanced Virgo\textquoteright{}s first three observing
  runs}},  {\em Phys. Rev. D} {\bf 104} (2021), no.~2 022005,
  [\href{http://arxiv.org/abs/2103.08520}{{\tt arXiv:2103.08520}}].

\bibitem{ValbusaDallArmi:2020ifo}
L.~Valbusa~Dall'Armi, A.~Ricciardone, N.~Bartolo, D.~Bertacca, and
  S.~Matarrese, {\it {Imprint of relativistic particles on the anisotropies of
  the stochastic gravitational-wave background}},  {\em Phys. Rev. D} {\bf 103}
  (2021), no.~2 023522, [\href{http://arxiv.org/abs/2007.01215}{{\tt
  arXiv:2007.01215}}].

\bibitem{Ananda:2006af}
K.~N. Ananda, C.~Clarkson, and D.~Wands, {\it {The Cosmological gravitational
  wave background from primordial density perturbations}},  {\em Phys. Rev. D}
  {\bf 75} (2007) 123518, [\href{http://arxiv.org/abs/gr-qc/0612013}{{\tt
  gr-qc/0612013}}].

\bibitem{Baumann:2007zm}
D.~Baumann, P.~J. Steinhardt, K.~Takahashi, and K.~Ichiki, {\it {Gravitational
  Wave Spectrum Induced by Primordial Scalar Perturbations}},  {\em Phys. Rev.
  D} {\bf 76} (2007) 084019, [\href{http://arxiv.org/abs/hep-th/0703290}{{\tt
  hep-th/0703290}}].

\bibitem{Konstandin:2011dr}
T.~Konstandin and G.~Servant, {\it {Cosmological Consequences of Nearly
  Conformal Dynamics at the TeV scale}},  {\em JCAP} {\bf 12} (2011) 009,
  [\href{http://arxiv.org/abs/1104.4791}{{\tt arXiv:1104.4791}}].

\bibitem{Agashe:2019lhy}
K.~Agashe, P.~Du, M.~Ekhterachian, S.~Kumar, and R.~Sundrum, {\it {Cosmological
  Phase Transition of Spontaneous Confinement}},  {\em JHEP} {\bf 05} (2020)
  086, [\href{http://arxiv.org/abs/1910.06238}{{\tt arXiv:1910.06238}}].

\bibitem{Ellis:2020nnr}
J.~Ellis, M.~Lewicki, and V.~Vaskonen, {\it {Updated predictions for
  gravitational waves produced in a strongly supercooled phase transition}},
  {\em JCAP} {\bf 11} (2020) 020, [\href{http://arxiv.org/abs/2007.15586}{{\tt
  arXiv:2007.15586}}].

\bibitem{Huber:2008hg}
S.~J. Huber and T.~Konstandin, {\it {Gravitational Wave Production by
  Collisions: More Bubbles}},  {\em JCAP} {\bf 09} (2008) 022,
  [\href{http://arxiv.org/abs/0806.1828}{{\tt arXiv:0806.1828}}].

\bibitem{Planck:2018vyg}
{\bf Planck} Collaboration, N.~Aghanim et~al., {\it {Planck 2018 results. VI.
  Cosmological parameters}},  {\em Astron. Astrophys.} {\bf 641} (2020) A6,
  [\href{http://arxiv.org/abs/1807.06209}{{\tt arXiv:1807.06209}}]. [Erratum:
  Astron.Astrophys. 652, C4 (2021)].

\bibitem{Linde:1996gt}
A.~D. Linde and V.~F. Mukhanov, {\it {Nongaussian isocurvature perturbations
  from inflation}},  {\em Phys. Rev. D} {\bf 56} (1997) R535--R539,
  [\href{http://arxiv.org/abs/astro-ph/9610219}{{\tt astro-ph/9610219}}].

\bibitem{Lyth:2001nq}
D.~H. Lyth and D.~Wands, {\it {Generating the curvature perturbation without an
  inflaton}},  {\em Phys. Lett. B} {\bf 524} (2002) 5--14,
  [\href{http://arxiv.org/abs/hep-ph/0110002}{{\tt hep-ph/0110002}}].

\bibitem{Enqvist:2001zp}
K.~Enqvist and M.~S. Sloth, {\it {Adiabatic CMB perturbations in pre - big bang
  string cosmology}},  {\em Nucl. Phys. B} {\bf 626} (2002) 395--409,
  [\href{http://arxiv.org/abs/hep-ph/0109214}{{\tt hep-ph/0109214}}].

\bibitem{Moroi:2001ct}
T.~Moroi and T.~Takahashi, {\it {Effects of cosmological moduli fields on
  cosmic microwave background}},  {\em Phys. Lett. B} {\bf 522} (2001)
  215--221, [\href{http://arxiv.org/abs/hep-ph/0110096}{{\tt hep-ph/0110096}}].
  [Erratum: Phys.Lett.B 539, 303--303 (2002)].

\bibitem{LIGOScientific:2016fpe}
{\bf LIGO Scientific, Virgo} Collaboration, B.~P. Abbott et~al., {\it
  {GW150914: Implications for the stochastic gravitational wave background from
  binary black holes}},  {\em Phys. Rev. Lett.} {\bf 116} (2016), no.~13
  131102, [\href{http://arxiv.org/abs/1602.03847}{{\tt arXiv:1602.03847}}].

\bibitem{Mandic:2016lcn}
V.~Mandic, S.~Bird, and I.~Cholis, {\it {Stochastic Gravitational-Wave
  Background due to Primordial Binary Black Hole Mergers}},  {\em Phys. Rev.
  Lett.} {\bf 117} (2016), no.~20 201102,
  [\href{http://arxiv.org/abs/1608.06699}{{\tt arXiv:1608.06699}}].

\bibitem{Dvorkin:2016okx}
I.~Dvorkin, J.-P. Uzan, E.~Vangioni, and J.~Silk, {\it {Synthetic model of the
  gravitational wave background from evolving binary compact objects}},  {\em
  Phys. Rev. D} {\bf 94} (2016), no.~10 103011,
  [\href{http://arxiv.org/abs/1607.06818}{{\tt arXiv:1607.06818}}].

\bibitem{Nakazato:2016nkj}
K.~Nakazato, Y.~Niino, and N.~Sago, {\it {Gravitational-Wave Background from
  Binary Mergers and Metallicity Evolution of Galaxies}},  {\em Astrophys. J.}
  {\bf 832} (2016), no.~2 146, [\href{http://arxiv.org/abs/1605.02146}{{\tt
  arXiv:1605.02146}}].

\bibitem{Dvorkin:2016wac}
I.~Dvorkin, E.~Vangioni, J.~Silk, J.-P. Uzan, and K.~A. Olive, {\it
  {Metallicity-constrained merger rates of binary black holes and the
  stochastic gravitational wave background}},  {\em Mon. Not. Roy. Astron.
  Soc.} {\bf 461} (2016), no.~4 3877--3885,
  [\href{http://arxiv.org/abs/1604.04288}{{\tt arXiv:1604.04288}}].

\bibitem{Jenkins:2019uzp}
A.~C. Jenkins and M.~Sakellariadou, {\it {Shot noise in the astrophysical
  gravitational-wave background}},  {\em Phys. Rev. D} {\bf 100} (2019), no.~6
  063508, [\href{http://arxiv.org/abs/1902.07719}{{\tt arXiv:1902.07719}}].

\bibitem{Alonso:2020mva}
D.~Alonso, G.~Cusin, P.~G. Ferreira, and C.~Pitrou, {\it {Detecting the
  anisotropic astrophysical gravitational wave background in the presence of
  shot noise through cross-correlations}},  {\em Phys. Rev. D} {\bf 102}
  (2020), no.~2 023002, [\href{http://arxiv.org/abs/2002.02888}{{\tt
  arXiv:2002.02888}}].

\bibitem{Adams:2013qma}
M.~R. Adams and N.~J. Cornish, {\it {Detecting a Stochastic Gravitational Wave
  Background in the presence of a Galactic Foreground and Instrument Noise}},
  {\em Phys. Rev. D} {\bf 89} (2014), no.~2 022001,
  [\href{http://arxiv.org/abs/1307.4116}{{\tt arXiv:1307.4116}}].

\bibitem{Korol:2021pun}
V.~Korol, N.~Hallakoun, S.~Toonen, and N.~Karnesis, {\it {Observationally
  driven Galactic double white dwarf population for LISA}},  {\em Mon. Not.
  Roy. Astron. Soc.} {\bf 511} (2022), no.~4 5936--5947,
  [\href{http://arxiv.org/abs/2109.10972}{{\tt arXiv:2109.10972}}].

\bibitem{Boileau:2020rpg}
G.~Boileau, N.~Christensen, R.~Meyer, and N.~J. Cornish, {\it {Spectral
  separation of the stochastic gravitational-wave background for LISA:
  Observing both cosmological and astrophysical backgrounds}},  {\em Phys. Rev.
  D} {\bf 103} (2021), no.~10 103529,
  [\href{http://arxiv.org/abs/2011.05055}{{\tt arXiv:2011.05055}}].

\bibitem{Thrane_2009}
E.~Thrane, S.~Ballmer, J.~D. Romano, S.~Mitra, D.~Talukder, S.~Bose, and
  V.~Mandic, {\it Probing the anisotropies of a stochastic gravitational-wave
  background using a network of ground-based laser interferometers},  {\em
  Physical Review D} {\bf 80} (dec, 2009).

\bibitem{Amaro-Seoane:2007osp}
P.~Amaro-Seoane, J.~R. Gair, M.~Freitag, M.~Coleman~Miller, I.~Mandel, C.~J.
  Cutler, and S.~Babak, {\it {Astrophysics, detection and science applications
  of intermediate- and extreme mass-ratio inspirals}},  {\em Class. Quant.
  Grav.} {\bf 24} (2007) R113--R169,
  [\href{http://arxiv.org/abs/astro-ph/0703495}{{\tt astro-ph/0703495}}].

\bibitem{Amaro-Seoane:2018gbb}
P.~Amaro-Seoane, {\it {Detecting Intermediate-Mass Ratio Inspirals From The
  Ground And Space}},  {\em Phys. Rev. D} {\bf 98} (2018), no.~6 063018,
  [\href{http://arxiv.org/abs/1807.03824}{{\tt arXiv:1807.03824}}].

\bibitem{LIGOScientific:2020iuh}
{\bf LIGO Scientific, Virgo} Collaboration, R.~Abbott et~al., {\it {GW190521: A
  Binary Black Hole Merger with a Total Mass of $150 M_{\odot}$}},  {\em Phys.
  Rev. Lett.} {\bf 125} (2020), no.~10 101102,
  [\href{http://arxiv.org/abs/2009.01075}{{\tt arXiv:2009.01075}}].

\bibitem{Babak:2017tow}
S.~Babak, J.~Gair, A.~Sesana, E.~Barausse, C.~F. Sopuerta, C.~P.~L. Berry,
  E.~Berti, P.~Amaro-Seoane, A.~Petiteau, and A.~Klein, {\it {Science with the
  space-based interferometer LISA. V: Extreme mass-ratio inspirals}},  {\em
  Phys. Rev. D} {\bf 95} (2017), no.~10 103012,
  [\href{http://arxiv.org/abs/1703.09722}{{\tt arXiv:1703.09722}}].

\bibitem{Bonetti:2020jku}
M.~Bonetti and A.~Sesana, {\it {Gravitational wave background from extreme mass
  ratio inspirals}},  {\em Phys. Rev. D} {\bf 102} (2020), no.~10 103023,
  [\href{http://arxiv.org/abs/2007.14403}{{\tt arXiv:2007.14403}}].

\bibitem{Alonso:2020rar}
D.~Alonso, C.~R. Contaldi, G.~Cusin, P.~G. Ferreira, and A.~I. Renzini, {\it
  {Noise angular power spectrum of gravitational wave background experiments}},
   {\em Phys. Rev. D} {\bf 101} (2020), no.~12 124048,
  [\href{http://arxiv.org/abs/2005.03001}{{\tt arXiv:2005.03001}}].

\bibitem{CMB-S4:2016ple}
{\bf CMB-S4} Collaboration, K.~N. Abazajian et~al., {\it {CMB-S4 Science Book,
  First Edition}},  \href{http://arxiv.org/abs/1610.02743}{{\tt
  arXiv:1610.02743}}.

\bibitem{Dimastrogiovanni:2022eir}
E.~Dimastrogiovanni, M.~Fasiello, A.~Malhotra, and G.~Tasinato, {\it {Enhancing
  gravitational wave anisotropies with peaked scalar sources}},
  \href{http://arxiv.org/abs/2205.05644}{{\tt arXiv:2205.05644}}.

\bibitem{Baker:2019ync}
J.~Baker et~al., {\it {High angular resolution gravitational wave astronomy}},
  {\em Exper. Astron.} {\bf 51} (2021), no.~3 1441--1470,
  [\href{http://arxiv.org/abs/1908.11410}{{\tt arXiv:1908.11410}}].

\bibitem{Tegmark:2003ve}
M.~Tegmark, A.~de~Oliveira-Costa, and A.~Hamilton, {\it {A high resolution
  foreground cleaned CMB map from WMAP}},  {\em Phys. Rev. D} {\bf 68} (2003)
  123523, [\href{http://arxiv.org/abs/astro-ph/0302496}{{\tt
  astro-ph/0302496}}].

\bibitem{Ichiki:2014qga}
K.~Ichiki, {\it {CMB foreground: A concise review}},  {\em PTEP} {\bf 2014}
  (2014), no.~6 06B109.

\bibitem{Paoletti:2022anb}
D.~Paoletti, F.~Finelli, J.~Valiviita, and M.~Hazumi, {\it {Planck and
  BICEP/Keck Array 2018 constraints on primordial gravitational waves and
  perspectives for future B-mode polarization measurements}},
  \href{http://arxiv.org/abs/2208.10482}{{\tt arXiv:2208.10482}}.

\bibitem{Graham:2017lmg}
P.~W. Graham and S.~Jung, {\it {Localizing Gravitational Wave Sources with
  Single-Baseline Atom Interferometers}},  {\em Phys. Rev. D} {\bf 97} (2018),
  no.~2 024052, [\href{http://arxiv.org/abs/1710.03269}{{\tt
  arXiv:1710.03269}}].

\bibitem{AEDGE:2019nxb}
{\bf AEDGE} Collaboration, Y.~A. El-Neaj et~al., {\it {AEDGE: Atomic Experiment
  for Dark Matter and Gravity Exploration in Space}},  {\em EPJ Quant.
  Technol.} {\bf 7} (2020) 6, [\href{http://arxiv.org/abs/1908.00802}{{\tt
  arXiv:1908.00802}}].

\bibitem{Kawamura:2020pcg}
S.~Kawamura et~al., {\it {Current status of space gravitational wave antenna
  DECIGO and B-DECIGO}},  {\em PTEP} {\bf 2021} (2021), no.~5 05A105,
  [\href{http://arxiv.org/abs/2006.13545}{{\tt arXiv:2006.13545}}].

\bibitem{Zhou:2023rop}
K.~Zhou, J.~Cheng, and L.~Ren, {\it {Detecting anisotropies of the stochastic
  gravitational wave background with TianQin}},
  \href{http://arxiv.org/abs/2306.14439}{{\tt arXiv:2306.14439}}.

\bibitem{Sato-Polito:2023spo}
G.~Sato-Polito and M.~Kamionkowski, {\it {Exploring the spectrum of stochastic
  gravitational-wave anisotropies with pulsar timing arrays}},
  \href{http://arxiv.org/abs/2305.05690}{{\tt arXiv:2305.05690}}.

\bibitem{Lasky:2015lej}
P.~D. Lasky et~al., {\it {Gravitational-wave cosmology across 29 decades in
  frequency}},  {\em Phys. Rev. X} {\bf 6} (2016), no.~1 011035,
  [\href{http://arxiv.org/abs/1511.05994}{{\tt arXiv:1511.05994}}].

\end{thebibliography}\endgroup
\end{document}